\documentclass{elsarticle}
\usepackage[a4paper, top=2.5cm, bottom=2.5cm, left=2.5cm, right=2.5cm]{geometry}
\usepackage{natbib}
\usepackage{longtable}
\setcitestyle{authoryear,round}
\makeatletter
\def\ps@pprintTitle{%
 \let\@oddhead\@empty
 \let\@evenhead\@empty
 \def\@oddfoot{}%
 \let\@evenfoot\@oddfoot}
\makeatother
\usepackage{amsmath}
    \usepackage{tikz}
\usepackage{ifthen}
\usepackage{graphicx}
\usepackage{enumitem}
\usepackage{ragged2e}
\usepackage{float}
\usepackage{todonotes}
\usepackage{placeins}
\usepackage{bigints}
\usepackage{pifont}
\usepackage{booktabs}

\DeclareMathOperator*{\argmin}{arg\,min}
\usepackage{lscape} 
\allowdisplaybreaks

\DeclareMathOperator*{\plim}{plim}

\usepackage{tabularx}
\usepackage{arydshln}
\usepackage{mathtools}
\usepackage{afterpage}
\usepackage{amsfonts}
\usepackage{algorithm,algpseudocode}
\usepackage{arydshln}
\usepackage[scr]{rsfso}
\usepackage{multirow}
\usepackage[flushleft]{threeparttable}
\usepackage{siunitx}
\usepackage{caption}
\usepackage{rotating}
\usepackage{amsfonts}
\usepackage{booktabs}
\usepackage{siunitx}
\usepackage{siunitx}
\usepackage{subcaption}
\usepackage{appendix}
\usepackage{amsthm}
\usepackage{amssymb,bm}
\usepackage{amsfonts}
\definecolor{applegreen}{rgb}{0.55, 0.71, 0.0}
\definecolor{ao(english)}{rgb}{0.0, 0.5, 0.0}
\usepackage{booktabs,caption}
\usepackage[pdftex,colorlinks,
  citecolor=black]{hyperref}
\usepackage{setspace}
\usepackage{bbold}

\newcommand{\norm}[1]{{\left\lVert#1\right\rVert}}

\usepackage{dsfont}
\usepackage{comment}
\usepackage{color}
\usepackage{graphicx}

\usepackage[normalem]{ulem}

\newcommand{\BS}[1]{\boldsymbol{#1}}
\newcommand{\add}[1]{{\textcolor{purple}{#1}}}

\newcommand{\bA}{\bm A}

\newcommand{\bB}{\bm B}

\newcommand{\bD}{\bm D}

\newcommand{\be}{\bm e}

\newcommand{\bof}{\bm f}

\newcommand{\bH}{\bm H}

\newcommand{\bI}{\bm I}

\newcommand{\bU}{\bm U}
\newcommand{\bu}{\bm u}

\newcommand{\bv}{\bm v}

\newcommand{\bw}{\bm w}

\newcommand{\bx}{\bm x}

\newcommand{\xmark}{\ding{55}}%

\newcommand{\bbeta}{\bm \beta}

\newcommand{\bdelta}{\bm \delta}

\newcommand{\boeta}{\bm \eta}

\newcommand{\bxi}{\bm \xi}

\newcommand{\gt}{g(N,T,\zeta)}

\newcommand{\bchi}{\bm \chi}

\newcommand{\bLambda}{\bm \varLambda}

\newcommand{\bSigma}{\bm \varSigma}

\newcommand{\bPsi}{\bm \varPsi}
\newcommand{\bOmega}{\bm \varOmega}
\newcommand{\var}{\mbox{Var}}
\newcommand{\cov}{\mbox{Cov}}

\newcommand{\eps}{\varepsilon}
\newcommand{\R}{\mathds{R}}

\newcommand{\A}{\boldsymbol{\mathfrak{A}}}

\newcommand{\Z}{\mathds{Z}}

\newcommand{\THRarg}[1]{\operatorname{THR}_{#1}}

\theoremstyle{definition}

\newtheorem{assumption}{Assumption}

\newtheorem{remark}{Remark}
\newtheorem{theorem}{Theorem}
\newtheorem{lemma}{Lemma}

\theoremstyle{plain}
\makeatletter
\def\adl@drawiv#1#2#3{%
        \hskip.5\tabcolsep
        \xleaders#3{#2.5\@tempdimb #1{1}#2.5\@tempdimb}%
                #2\z@ plus1fil minus1fil\relax
        \hskip.5\tabcolsep}
\newcommand{\cdashlinelr}[1]{%
  \noalign{\vskip\aboverulesep
           \global\let\@dashdrawstore\adl@draw
           \global\let\adl@draw\adl@drawiv}
  \cdashline{#1}
  \noalign{\global\let\adl@draw\@dashdrawstore
           \vskip\belowrulesep}}
\makeatother



\begin{document}
\bibliographystyle{apalike2}

\begin{frontmatter}
\title{\textcolor{black}{Decomposing Global Bank Network Connectedness:\\ 
What is Common, Idiosyncratic and When?}}
\author[add1]{Jonas Krampe}
\ead{jonas.krampe@cornell.edu}
\author[add2]{Luca Margaritella\corref{cor2}}
\ead{luca.margaritella@nek.lu.se} 
\address[add1]{Cornell University, Department of Statistics and Data Science}
\address[add2]{Lund University, Department of Economics}
\cortext[cor2]{Corresponding author. We wish to thank: Matteo Barigozzi, Simon Reese, Rosnel Sessinou, Joakim Westerlund for the insightful discussions that helped shaping and improving the paper. All remaining errors are our own. The authors disclose no conflicts of interest.}
\date
\maketitle
\begin{abstract}
We propose a novel approach to estimate high-dimensional global bank network connectedness in both the time and frequency domains. By employing a factor model with sparse VAR idiosyncratic components, we decompose system-wide connectedness (SWC) into two key drivers: (i) common component shocks and (ii) idiosyncratic shocks. We also provide bootstrap confidence bands for all SWC measures. Furthermore, spectral density estimation allows us to disentangle SWC into short-, medium-, and long-term frequency responses to these shocks.
We apply our methodology to two datasets of daily stock price volatilities for over 90 global banks, spanning the periods 2003-2013 and 2014-2023. Our empirical analysis reveals that SWC spikes during global crises, primarily driven by common component shocks and their short-term effects. Conversely, in normal times, SWC is largely influenced by idiosyncratic shocks and medium-term dynamics.

\bigbreak
\end{abstract}
\end{frontmatter}

\section{Introduction}
Measuring \emph{connectedness} is of paramount importance in many aspects of financial risk measurement and management. Particularly, following the global financial crisis in 2008–2009, the heightened focus of governments and financial institutions on the significant concerns surrounding the propagation of macro financial risks and its potential impact on financial stability has become increasingly evident. Connectedness measures, such as return connectedness, default connectedness, and system-wide connectedness, are commonly featured in various facets of risk management, including market risk, credit risk, and systemic risk.
Nevertheless, the concept of connectedness remained rather elusive in econometric theory until \citet{diebold2014network} undertook the task of addressing it comprehensively. 
Their work provided a rigorous definition by introducing measures of connectedness rooted in (generalized) forecast error variance decomposition (FEVD) from approximating, finite order vector autoregressive (VAR) models.\footnote{To clarify: the term \emph{approximating} refers to the fact that a model should be chosen for the data, and that is never correct; if a dynamic one is chosen then, like a VAR here, a finite length of its past dynamic (i.e., a lag-length) has to be specified. This in itself is another approximation, as it presumes all the series to have the same dynamic.}  To elaborate, their approach involves evaluating the distribution of forecast error variance across different actors, such as banks, firms, markets, countries, etc., attributable to shocks originating elsewhere. In simpler terms: if the future variation of e.g., bank $i$, is mostly due to shocks attributable to bank $j$, then the two banks are connected as $j\to i$, and vice versa. 
Then, the appeal of such approach lies in its ability to address the question of ``to what extent the future variation (at different horizons `$H$') of actor $i$ 
can be attributed \emph{not} to internal shocks originating within actor $i$ itself, but rather to external factors associated with actor $j$?". 
To identify uncorrelated structural shocks from correlated reduced form shocks, \citet{diebold2014network} chose the generalized variance decomposition (GVD) framework introduced in \citet{koop1996impulse, pesaran1998generalized}. Differently from the identification schemes that orthogonalize the shocks e.g., through Cholesky factorization, and which are dependent on the variables' ordering, GVD avoids forced orthogonalization of the shocks and -under a normality assumption- properly accounts for historically observed correlations among them, while being order-invariant.\footnote{The principle of GVD and their generalized impulse responses is that of treating each variable as if they were the first in the ordered vector of observables, and account for the correlation among shocks by discounting for historical correlations among them, rather than orthogonalize them. In this sense, it does not matter which variable comes first or later in the vector.} \par Although it clearly depends on the level of aggregation considered, systems of banks, firms, markets, countries etc.~are seldom low-dimensional. The likely high-dimensionality of such systems instead introduces some challenges to tackle in order to estimate the approximating --now high-dimensional-- models, imposed on the --now high-dimensional-- vector of observables. 
Such challenges have been taken on by \citet{demirer2018estimating} in the context of global bank network connectedness. Using a \emph{sparse} VAR model of order $p$, VAR($p$), directly on the observables, they employ $\ell_1 + \ell_2$-norm regularization in the form of an (adaptive) Elastic Net. This approach allows them to jointly perform shrinkage, variable selection, and estimation where their goal is that to estimate the high-dimensional connectedness network linking a publicly traded subset of the world's top 150 banks, covering the period from 2003 to 2013. Once an estimate of the high-dimensional VAR coefficient matrix is obtained, the $H$-step generalized variance decomposition matrix can be easily computed and thus the various connectedness relationships as in \citet{diebold2014network}. These can be between: each pair of banks (\emph{pairwise directional connectedness}), each bank with \emph{all} the others -and vice versa- 
(\emph{total directional connectedness}), and all banks in a total connectedness sense 
(\emph{system-wide connectedness}, SWC henceforth). We refer to Section \ref{sec_model} for the mathematical definitions.\par As any measure based on a model relies on a set of decisions/assumptions on that very model, the connectedness measure of \citet{demirer2018estimating} is no exception. As observed in \citet{diebold2014network}, among other factors, estimating connectedness based on FEVD is affected by the type of approximating model to which data is fed to, and forecast error variance is obtained from. The popularized use of sparse-regularization techniques to account for the large dimensionality of such problems can be tempting, if anything for its simplicity, and -aside of \citet{demirer2018estimating}- it is often employed in the applied literature on financial connectedness \citep[see a.o.,][]{yi2018volatility, liu2022high}.  
\par In this paper, we argue that such a direct \emph{sparsity} assumption on the VAR coefficient matrix might be a (too) strong statement on the data generating process, and becomes (more) reasonable only \emph{after} controlling for common variation within the observables, i.e., after estimation and thus accountancy of the common factors in the data. In fact, common factors are widely recognized to play a fundamental role with financial data and its modeling.\footnote{As observed in \citet{bai2006evaluating}, e.g., the arbitrage pricing theory is built upon the existence of a set of common factors underlying all asset returns. 
In the capital asset pricing theory the market return is the common risk factor that has pervasive effects on all assets. Many other examples could be made.} 
Failure to account for factors, and the use of direct sparsity assumptions when linkages among units are truly non-sparse might induce an underestimation of the degree of connectedness.  However, here we do not depart away from high-dimensional \emph{sparse} VARs, but instead build upon the recent literature bridging factors and sparse models \citep{fan2021bridging, barigozzi2023fnets, krampe2021factor}, where we assume the series to follow an approximate, static factor model whose idiosyncratic term follows a \emph{sparse} VAR.\footnote{We allow both factors and idiosyncratics to have parametric VAR representations. The term ``static" \citep[see][]{stock2002forecasting, bai2003} is to distinguish it from a model where lagged factors enter directly the factor model decomposition: the so-called 'generalized dynamic' factor model \citep{forni2000generalized}. The term 'approximate' refer to the fact that the idiosyncratics are allowed to exhibit cross-sectional dependence. We refer to Section \ref{sec_techdet} for the details.} We also employ the same strategy in the frequency domain in Section \ref{sec:spectral}, where we describe the frequency dynamics of the connectedness by considering the spectral representation of variance decompositions based on the frequency responses to
shocks.
\par \emph{Why a factor model for computing connectedness?} As mentioned, factor models play a fundamental role in financial data analysis, as documented in a nowadays vast literature. Assuming sparsity directly on the coefficient matrix of the VAR is tantamount to force somewhat weaker predictive linkages to be zeroed-out by the LASSO-type technique employed. While regularization promotes parsimony (interpretability) and contrasts overfitting, in the case of connectedness it risks to underestimate the degree of SWC by tossing away connections. A factor model instead, accounts for a common dynamic among all the volatilities. Once that has been accounted for, the idiosyncratic dynamic of connections left is much more sensible to be sparse. Also, we propose here a \emph{joint} treatment of factors and idiosyncratics (not either of). That means our FEVD expression (see \eqref{theta} below) contains both moving average (MA) representations of factors \emph{and} idiosyncratics. As a consequence, we can compute high-dimensional IRFs and therefore the connectedness measures proposed in \citet{diebold2014network}, 
now disentangled between common and idiosyncratic shocks.  
Especially, SWC can be divided into SWC \emph{due to the common shocks} and SWC \emph{due to the idiosyncratic shocks}. This helps in addressing questions such as: ``what drives SWC in the banking sector?" and also, ``is a shock on a single bank (and likewise a global shock) able to (and to what extent) affect the SWC? and when?".
\par \emph{Why factors \& idiosyncratics?} First, explicit modeling of the idiosyncratics allows to capture cross-sectional and time dependence, which remains after the factors' estimation. If instead, what remains is only measurement error, this is unnecessary. But while this scenario might be defensible in macroeconomic contexts, it is really not the case in finance \citep[see e.g, ][]{acemoglu2012network}. Thinking about stock returns daily range-based volatilities for a publicly traded subset of the world’s top 150 banks, as in \citet{demirer2018estimating}, it is reasonable to assume a common dynamic among these banks' stock price volatilities, i.e., some sort of ``market dynamic". Likewise, it is also sensible that a substantial ``individual dynamic" of the single banks themselves, or small subsets of them, would play a role. Second, once the common factors are accounted for, the assumption of \emph{sparsity}—which is often considered unrealistic on its own (e.g., \citealp{giannone2021economic})—becomes far more plausible when imposed on the idiosyncratic VAR coefficient matrices.\footnote{Note how in the literature there exists many papers \citep[a.o.,][]{billio2012econometric, hecq2023granger} taking on the challenge of estimating financial networks (not necessarily \emph{connectedness} networks) via direct regularization of high-dimensional VARs, as also done in \citet{demirer2018estimating}. As sparsity is a non-testable assumption, assuming it directly on the VAR coefficient matrix can be, at times, hard to justify.} Third, controlling for common factors in a first step tends to reduce collinearity among idiosyncratics, which is well known to render LASSO variable selection arbitrary. Fourth, the factor model gets robustified against misspecifications of the number of factors, since the transferred mistake to the idiosyncratics is at least modeled, instead of ignored. About this last point, this type of modeling also attenuates potential worries about rate-weak factors going undetected.
\citet{barigozzi2024dynamic} highlight that factors might remain undetected when their empirical cross-correlations are small. Regardless of whether they are static or dynamic, such weak factors are not lost but instead remain within the empirical idiosyncratic space. Due to their limited correlations, this omission is generally inconsequential—provided that the idiosyncratic component itself is not disregarded.

 \par To illustrate our approach, we begin by re-examining the estimation of global bank connectedness networks using the dataset from \citet{demirer2018estimating}. This contains stock price volatilities for a large set of global banks, as well as the bond price volatilities of ten major countries, recorded on daily frequency from 2003 until 2014. Next, we construct a more recent dataset covering 2014–2023, which includes nearly the same bank and bond assets, and replicate our analysis. This updated vintage allows us to analyze how SWC has evolved from 2014 until more recently, but also to investigate similarities and differences in the behavior of connectedness across different global crises (e.g., Covid19).  
 We employ an \emph{approximate static factor model} with sparse VAR idiosyncratics just like the one considered in \citet{krampe2021factor}. The common factors and loadings are estimated via principal component analysis (PCA), while the (obtained) idiosyncratics are estimated in a sparse VAR by adaptive LASSO. The compound of the two estimates in moving average representation gives the response of the observables, and the sequence of moving average coefficients at different horizons $H$ gives the impulse response of the observables to either a global or an idiosyncratic shock. Consequently, forecast error variance decompositions can be obtained and likewise a measure of SWC, now declined into common and idiosyncratic shocks. Additionally, adapting the framework of \citet{krampe2023structural} we are able to compute bootstrap confidence bands for the SWC. This is an important addition, as previously no statistical error bands were given over the estimated connectedness.
 \par We also extend the analysis to the spectral domain. This entails estimating the spectral densities of factors and idiosyncratics, the first with traditional nonparametric methods, the second leveraging on the VAR structure and using regularization to estimate its (high-dimensional) residuals' precision matrix. 
 \par Let us note that one could also directly estimate a sparse moving average representation of the idiosyncratics using a high-dimensional version of the local projection (LP) \citep[][]{jorda2005estimation}. However, even in the low dimensional case ``there is no evidence to suggest that local LPs should replace conventional linear VAR models" \citep[][]{kilian2017structural}.
 \par Our empirical findings demonstrate how in calm times SWC is high, but mostly due to idiosyncratic variation at low frequency. When financial turmoils occur, SWC is even higher, and the common component variation spikes upward, driven by a short-run dynamic response to shocks. This is interesting, as it blends with -and contribute to- the economic literature discourse on systemic risk and stability in financial networks. \citet{acemoglu2015systemic} have shown how there exists a ``double-edge knife" component to connectedness in financial networks. On the one hand, highly interconnected financial networks are ``shock-absorbing". On the other hand, once a certain unspecified threshold of connectedness is passed, the robustness turns into a ``shock propagating" mechanism. What we find is essentially that networks are shock-absorbing as long as their connectedness is driven by an idiosyncratic dynamic. Networks are not anymore shock-absorbing as soon as connectedness starts to be driven more by a common component dynamic at high frequency.

 \par Particularly in the context of systemic risk, measuring connectedness has been extensively explored in the literature, offering a variety of methods, each of which has its own advantages and limitations. Our focus is on showing that employing a factor model with sparse VARs idiosyncratics allows to answer a richer question within the context of estimation of global bank connectedness networks. Therefore, for comparison purposes we employ the same GVD-based identification as \citet{demirer2018estimating}.  
 
 Let us now mention few important related works.  \citet{barigozzi2017network} look at generalized dynamic factor models to study interdependences in large panels of financial series, specifically S\&P100. Connectedness networks for the idiosyncratics are built, based on FEVD, but they focus mainly on the idiosyncratics, \emph{after} controlling (filtering) for the ``market effects", i.e., after accounting for the factors. \par Similarly, \citet{ando2022quantile} employ a VAR together with a common factor error structure, fitted by quantile regression. Also in this case, their focus is on the analysis of direct spillovers of credit risk, \emph{after} controlling for common systematic factors. This means that their vector of forecast errors for the target is conditional on the information set (at time $t-1$) and, crucially, on the common factors. Their FEVD is then a measure of the proportion of the $h$-steps-ahead forecast error variance in the $j$-th observable, accounted for by the $i$-th idiosyncratic innovation.
 \par Another interesting related approach is that of \citet{barigozzi2021time} who introduce a time-varying general dynamic factor model for high-dimensional locally
stationary processes. Their focus though is on the factors only. Similar to our empirical findings, using a panel of adjusted intra-day (1999-2015) log ranges for 329 constituents of
the S\&P500, they show how large increases in connectedness (intended as \emph{factors-only} connectedness) are associated with the most important turmoils in the stock market (e.g., the great financial crisis of 2007–2009).
  
 \par The main difference between \citet{barigozzi2017network, ando2022quantile, barigozzi2021time} treatments and the one we propose is that  
  we consider a \emph{joint} factor-plus-idiosyncratic treatment of the IRFs in order to compute SWC based on FEVD. To elaborate, although we do not concentrate on the tails as in \citet{ando2022quantile}, nor on locally stationary processes as in \citet{barigozzi2021time},
  we allow both MA representations of factors \emph{and} idiosyncratics to enter the expression of the FEVD. Thus, connectedness due to factors, idiosyncratics and the summation of both can be properly disentangled, without limiting it to be computed only from either of these sources. Also, we work out a complete extension to the frequency domain, which allows to disentangle further the financial connectedness into long/medium/short term responses to shocks. This is relevant and, to the best of our knowledge, not considered before within this context.
\section{Connectedness measures, approximating factor model and estimation strategy}\label{sec_model}
In this section, we first briefly introduce the connectedness measures established by \citet{diebold2014network}. Then, we discuss how to adapt this framework when the employed model is an approximate static factor model with sparse VAR idiosyncratics.  We show how this modeling approach opens up to more flexibility in interpretation as it disentangles the connectedness due to the common component shocks, to that due to the idiosyncratic shocks. We first present the framework in-population, then briefly discuss its estimation strategy in-sample. As in Section \ref{sec_data} we are going to use this proposed approach on a couple of global banking datasets, throughout this section, whenever we talk about ``observables", we have in mind stock return daily range-based realized volatilities\footnote{Throughout, for brevity, we often omit the ``realized" and leave only ``volatility"; it should always be intended as ``realized volatility".} for a (large) set of world banks (details are given in Section \ref{sec_data}). \par In what follows, we employ boldface characters for vectors and capital boldface characters for matrices where e.g., $\bI_N$ is the identity matrix of order $N$. As the notation used in this section is always defined in-text, we refer the reader to the first paragraph of Section \ref{sec_techdet} on Technical Details, where a more detailed description of the notation is provided.

\par Consider a large, $N$-dimensional covariance-stationary stochastic process with MA representation $\bx_t=\sum_{i=0}^{\infty}\bPsi_i \bu_{t-i}$, $\bu_t\sim (\boldsymbol{0}, \bSigma)$, $\bPsi_0=\bI_N$.  
Then, bank $j$'s contribution to bank $i$'s $H$-steps ahead generalized error variance, i.e., \emph{pairwise directional connectedness} is given, in population, by\footnote{Note how since the GVD of \citet{koop1996impulse} is employed and therefore the variance shares are not guaranteed to add up to 1, each entry of the generalized variance decomposition matrix gets normalized by the row sum $\sum_{j=1}^N \theta_{ij}^g(H)$. This way, $\sum_{j=1}^N C_{i\leftarrow j}^H=1$ and $\sum_{i,j=1}^N C_{i\leftarrow j}^H=N$.} 
\begin{align}\label{conn_1}
    C_{i\leftarrow j}^H=\frac{\theta_{ij}^g(H)}{\sum_{j=1}^N\theta_{ij}^g(H)}, \quad \text{where:}\quad \theta_{ij}^g(H)=\frac{\sigma_{jj}^{-1}\sum_{h=0}^{H-1} (\BS e_i^{\top}\BS \Psi_H\BS \Sigma \BS e_j)^2}{\sum_{h=0}^{H-1}\BS e_i^{\top}\BS \Psi_H\BS \Sigma \BS \Psi_H^{\top}\BS e_i}, \quad H=1,2,\ldots ,
\end{align}\normalsize
and where $\BS e_i (\BS e_j)$ is a selection vector with $i (j)$-th element unity and zeros elsewhere and $\sigma_{jj}=\be_j^{\top}\bSigma \be_j$. 
There, $\theta_{ij}^g(H)$ for $i,j=1,\ldots,N$, is the forecast error variance decomposition, i.e., the  proportion of the $H$-step ahead forecast error variance of the volatility of stock price of bank $i$, accounted for by the innovations in the volatility of stock price of bank $j$.
Similarly, for the \emph{total directional connectedness} $C^H_{i\leftarrow \operatorname{All}(j)}$,  $C^H_{\operatorname{All}(j)\leftarrow i}$ and\footnote{We use the notation $\operatorname{All}(j)=\{i: i\neq j\}$, $\operatorname{All}(i)=\{j: j\neq i\}.$} \emph{system-wide connectedness} $C^H$:

\begin{align}\label{conn_2}
 C^H_{i\leftarrow \operatorname{All}(j)}=\frac{\sum_{\substack{j=1\\j\neq i}}^N C_{i\leftarrow j}^H}{N}; \quad  C^H_{\operatorname{All}(j)\leftarrow i}=\frac{\sum_{\substack{j=1\\j\neq i}}^N C_{j\leftarrow i}^H}{N}; \quad C^H= \frac{\sum_{\substack{i,j=1\\j\neq i}}^N C_{i\leftarrow j}^H}{N}.
\end{align}\normalsize
\par In this paper, in place of assuming a VAR($p$) approximation for $\bx_t$ as in \citet{demirer2018estimating}, we first assume that the $N$ time series can be decomposed into a sum of two uncorrelated components: an $N$-dimensional vector of common components $\bchi_t$, and an 
$N$-dimensional vector of idiosyncratic components $\bxi_t$, such that:  \begin{align} \label{eq_decom}
    \bx_t=\bchi_t+\bxi_t.
\end{align} \normalsize

As for the first, $\bchi_t$, it represents the comovements between the $N$ bank stock price volatilities, and it is assumed to be low-rank, i.e., driven linearly by an $r$-dimensional vector of common factors $\bof_t$, for $r\ll N$. This means there are $r$ factors, common to all the different banks, driving the change of their stock price volatilities. We call this common behavior ``\emph{the market}". Provided a consistent estimate of $\bof_t$ is obtained, and likewise one for $\bLambda$, i.e., the $N\times r$ matrix of factor loadings on $\bx_t$, this entails for the common component an effective dimensionality reduction from $N$ to $r$ series.\footnote{Shall be noted here that a factor model in itself is \emph{never} a dimensionality reduction technique. From $N$ observables to $2N$ with the decomposition. It \emph{is} a reduction if one assumes both a low rank for $\bchi_t$ and white noise for $\bxi_t$. The low rank assumption is mostly sensible, the white noise on $\bxi_t$ is often not.} Therefore, the common component gets decomposed as $\bchi_t=\bLambda \bof_t$. As for the idiosyncratic component, $\bxi_t$, it represents individual features of the series \textit{and/or} measurement error. E.g., certain stocks might be more exposed to the behavior of their own reference stock exchange, or to the political situation of their origin country, or to the monetary policy decision of the central bank of their origin country, or to the exchange rate risk, etc. For the purpose of forecasting $\bx_t$, if $\bxi_t$ would truly only be made up of measurement errors, its inclusion in the forecasting equation should not be relevant. However, if $\bxi_t$ contains individual features of the series, and these are correlated (e.g., two banks listed on the same stock exchange), accounting for idiosyncratics in the forecasting equation becomes paramount. \par Instead of assuming a stable VAR($p$) on $\bx_t$, we assume \emph{two} stable VARs, namely a VAR($p_f$) for $\bof_t$ and a VAR($p_{\xi}$) for $\bxi_t$, such that 
\begin{align}\label{eq_vars}
    \bof_t=\sum_{j=1}^{p_f}\BS D^{(j)} \BS f_{t-j}+\BS u_t, \quad \;\;\bxi_t=\sum_{j=1}^{p_{\xi}}\BS B^{(j)} \BS \xi_{t-j}+\BS v_t.
\end{align} \normalsize
Then, the factor model decomposition in \eqref{eq_decom} can be re-written as:
\begin{align}
\boldsymbol{x}_t&=\boldsymbol{\Lambda}\left(\sum_{j=1}^{p_f}\BS D^{(j)} \BS f_{t-j}+\BS u_t\right)+\sum_{j=1}^{p_{\xi}}\BS B^{(j)} \BS \xi_{t-j}+\BS v_t \notag\\
&=\boldsymbol{\Lambda}\sum_{j=0}^{\infty}\BS \Psi^{(j)}_f \BS u_{t-j}+\sum_{j=0}^{\infty}\BS \Psi^{(j)}_{\xi} \BS v_{t-j} =\sum_{j=0}^{\infty} \begin{pmatrix}
        \boldsymbol{\Lambda}\BS \Psi^{(j)}_f&\BS \Psi^{(j)}_{\xi}
    \end{pmatrix} \boeta_{t-j},\label{eq_irf}
\end{align}\normalsize where the second line rewrites the VARs in \eqref{eq_vars} for factors and idiosyncratics in their infinite moving average representations, for $\bPsi_f^{(0)}, \bPsi_{\xi}^{(0)}=\bI_r,\bI_N$, $\bPsi_f^{(j)}, \bPsi_{\xi}^{(j)}=\boldsymbol{0}$ if $j<0$ and $\boeta_t:=(\BS u_t^\top,\BS v_t^\top)^\top\overset{iid}{\sim}(\boldsymbol{0}, \BS \Sigma_{\boeta})$ such that $\BS \Sigma_{\eta}$ is an $(r+N)\times (r+N)$ block-diagonal matrix with blocks $\BS\Sigma_u$, $\BS \Sigma_v$, i.e., respectively the covariance matrices of factors and idiosyncratics innovations. Within this framework, an impulse response function (IRF) would measure the time profile of the effect of a \emph{market} and/or an \emph{idiosyncratic} shock at a given point in time on the expected future values of (any of) the observables in $\bx_t$. More formally, IRFs here compare the time profile of the effect of an hypothetical $r$-dimensional \emph{market}-shock $\bdelta^m=(\delta_1^m,\ldots,\delta_r^m)^{\top}$ and/or an $N$-dimensional \emph{idiosyncratic} shock $\bdelta^{id}=(\delta_1^{id},\ldots,\delta_N^{id})^{\top}$ hitting the global banking system at time $t$ (i.e., $\bu_t=\bdelta^m$ and/or $\bv_t=\bdelta^{id}$), with a base-line profile at time $t+H$, given (i.e., conditional on) the global banking system behavior's history up to before the shock, i.e., $\bOmega_{t-1}$. Letting $\bdelta=(\bdelta^{m \top}, \bdelta^{id \top})^{\top}$, then the IRF captures the following (expected) difference: $IRF(H,\bdelta,\bOmega_{t-1})=\mathbb{E}(\bx_{t+H}|\boeta_t=\bdelta, \bOmega_{t-1})-\mathbb{E}(\bx_{t+H}|\bOmega_{t-1})$, which translated into \eqref{eq_irf} means $IRF(H,\bdelta,\bOmega_{t-1})=\begin{pmatrix}
        \boldsymbol{\Lambda}\BS \Psi^{(H)}_f&\BS \Psi^{(H)}_{\xi}
    \end{pmatrix}\bdelta.$ The usual problem with this formulation is that while it is independent of $\bOmega_{t-1}$, the IRF depends on the composition of the vector $\bdelta$, i.e., the vector of hypothesised shocks. The \emph{generalized} IRF (GIRF) approach of \citet{koop1996impulse, pesaran1998generalized} adopted in \citet{diebold2014network} and by us as well, is that of avoiding orthogonalization of the shocks in $\boeta_t$, but instead using the expression for the IRF directly, shocking only one element (say, the $i$th) at a time, and integrating out the (expected) effects of the other shocks $\mathbb{E}(\boeta_t|\eta_{i,t}=\delta_i)$ via an assumed or (historically) observed distribution of the errors. Indeed, by assuming $\boeta_t$ to be multivariate Gaussian for instance, then by standard properties\footnote{For any two zero mean Gaussian random variables $Y,X$ with variance $\sigma_y, \sigma_x$ respectively, then $\mathbb{E}(Y|X=x)=\sigma_y \rho(x/\sigma_x)$. The normality assumption is mostly for convenience; as noted in \citet{pesaran1998generalized} one can obtain the conditional expectation $\mathbb{E}(\boeta_t|\eta_{i,t}=\delta_i)$ by stochastic simulations or resampling techniques.} one gets $\mathbb{E}(\boeta_t|\eta_{i,t}=\delta_i)=\bSigma_{\eta}\be_i (\be_i^{\top}\bSigma_{\eta}\be_i)^{-1}\delta_i$, where again  $\BS e_i$ is a selection vector with $i$-th element unity and zeros elsewhere. Then, by setting $\delta_i=(\be_i^{\top}\bSigma_{\eta}\be_i)^{1/2}$, GIRF and FEVD ($\theta^g_{ij}$) are obtained for $H=1,2,\ldots,$ as
          \begin{align}
         &\operatorname{GIRF}(H, \delta_j)=\frac{\begin{pmatrix}
        \boldsymbol{\Lambda}\BS \Psi^{(H)}_f \;\BS \Psi^{(H)}_{\xi}
    \end{pmatrix} \BS \Sigma_{\eta}\BS e_j}{\left(\BS e_j^{\top}\BS \Sigma_{\eta} \BS e_j\right)^{1/2}}, \quad j=1,\ldots, r+N,\label{girf}\\
&\theta_{ij}^g(H)=\frac{(\be_j^{\top}\bSigma_{\eta}\be_j)^{-1}\sum_{h=0}^{H-1}\left( \BS e_i^{\top}\begin{pmatrix}
        \boldsymbol{\Lambda}\BS \Psi^{(h)}_f \;\BS \Psi^{(h)}_{\xi}
    \end{pmatrix} \BS \Sigma_{\eta}\BS e_j\right)^2}{\left(\sum_{h=0}^{H-1}\BS e_i^{\top}\begin{pmatrix}
        \boldsymbol{\Lambda}\BS \Psi^{(h)}_f \;\BS \Psi^{(h)}_{\xi}
    \end{pmatrix} \BS \Sigma_{\eta} \begin{pmatrix}
        \boldsymbol{\Lambda}\BS \Psi^{(h)}_f \;\BS \Psi^{(h)}_{\xi}
    \end{pmatrix}^{\top}\BS e_i\right)}, \quad i=1,\ldots,N, \; j=1,\ldots, r+N.\label{theta}
     \end{align} 

  \normalsize   
     Clearly, both GIRF and the FEVD can be now split into a ``due to a market shock" and ``due to an idiosyncratic shock". This is obtained simply by, respectively, either {specifying $j=1,\ldots,r$ in \eqref{girf}, \eqref{theta} for ``market only", and $j=r+1,\ldots, r+N$ in \eqref{girf}, \eqref{theta}, for ``idiosyncratics only".}  
     As a consequence, the same connectedness measures as in \eqref{conn_1}, \eqref{conn_2} can be obtained, now decomposed into: (\emph{pairwise directional, total), system-wide} connectedness due to a market shock, $C^H_{Mkt}$, due to an idiosyncratic shock, $C^H_{Ids}$, and due to the summation of both $C^H=C^H_{Mkt}+C^H_{Ids}$.
     \begin{align}
         C^H_{Mkt}= \frac{\sum_{i=1}^N\sum_{\substack{j=1}}^r C_{i\leftarrow j}^H}{N}, \quad C^H_{Ids}= \frac{\sum_{i=1}^N\sum_{\substack{j=r+1\\j-r\neq i}}^{r+N} C_{i\leftarrow j}^H}{N}, \quad C^H= C^H_{Mkt}+C^H_{Ids},
     \end{align}\normalsize
where $C_{i\leftarrow j}^H$ are the same as defined in \eqref{conn_1}, but now containing $\theta_{ij}^g(H)$ as in \eqref{theta} 
\subsection{Estimation}
\par All we presented so far was in-population. In order to obtain an in-sample estimate of \eqref{eq_irf}, a two step procedure as in \citet{krampe2021factor} is employed here, that estimates the factor(s) and loadings first, and the sparse VAR over the idiosyncratics after. We leave the technical details/assumptions for Section \ref{sec_techdet}, but the estimation steps and the intuition of how this work in relation to \eqref{eq_irf} is now given. 
\begin{itemize}
    \item[(I)] Factors $\bof_t$ and loadings $\bLambda$ are estimated via PCA. Then, a VAR($p_f$) is fit via least squares on the estimated factors $\hat{\bof}_t$, in order to retrieve the estimates of the autoregressive parameters ($\bD^{(j)}, j=1,\ldots,p_f$ in \eqref{eq_irf}).
    \item[(II)] From (I), the $N$-dimensional vector of VAR residuals $\hat{\bxi}_t=\bx_t-\hat{\bLambda}\sum_{j=1}^{p_f}\hat{\bD}^{(j)}\hat{\bof}_{t-j}$ is then the (high-dimensional) vector of (estimated\footnote{The sparse VAR over the idiosyncratics is effectively an estimation of a pre-estimated quantity. \citet{krampe2021factor} derive and bound the expression of the estimation error coming from the first step where factors are estimated.}) idiosyncratic components upon which a sparse VAR($p_{\xi}$) via adaptive LASSO is fit. Letting $\hat {\BS{\xi}}_t^v=(\hat {\BS\xi}_t^\top,\dots,\hat {\BS\xi}_{t-p_{\xi}}^\top)^\top$, then an adaptive LASSO estimator for $\bbeta^{(j)}$ i.e., the $j$th row of $(\bB^{(1)},\dots,\allowbreak \bB^{(p_{\xi})}),$ is given by 
\begin{align}
    \label{eq.LASSO.beta.j}
\hat{\bbeta}^{(j)}=\argmin_{\bbeta \in \R^{Np_{\xi}}} \frac{1}{T-p_{\xi}} \sum_{t=p_{\xi}+1}^T (\hat \xi_{j,t}-\bbeta^\top \hat {\BS\xi}_{t-1}^v)^2+\lambda \sum_{i=1}^{Np_{\xi}} |g_i \beta_i|,\quad j=1,\dots,N, 
\end{align}\normalsize where $\lambda$ is a tuning parameter determining the strength of the shrinkage and $g_i$ are first-step LASSO weights (see Section \ref{sec_techdet} for details, including choice of $p_f, p_{\xi}$).
\end{itemize}

By inverting the estimated VARs for factors and idiosyncratics as of (I) and (II), in their moving average representations, we then obtain $\hat \bPsi_{f}$, $\hat \bPsi_{\xi}$ as of second line of \eqref{eq_irf}. Note that the inversion is simply an algebraic nonlinear transformation which can result in a sparse VAR being represented as a nonsparse MA.
Finally, the covariance matrix of the error $\bSigma_{\eta}$ is estimated by plugging-in the upper left block of the sample covariance of the residuals from the VAR($p_f$) for the factors, $\hat{\bSigma}_u$, and on the bottom right block the sample covariance of the residuals from the regularized VAR($p_{\xi}$) for the idiosyncratics, $\hat{\bSigma}_v$. \par Importantly, we show that it is also possible to obtain confidence bands 
for the connectedness measures.
The algorithm for computing the bootstrap confidence bands is given in full in Section \ref{sec_boot}. 
The statistical validity of the employed bootstrap is given by results in \citet{krampe2023structural}. 
\par Later, in our empirical application in Section \ref{sec_data} we are going to focus especially on $C^H$, the SWC, as it is the most interesting in a systemic-risk perspective. \citet{demirer2018estimating} found that SWC has grown steadily between $2004$ and $2008$, peaking with the financial crisis, only to then decrease again (although not recovering the initial level) all the way to $2013$. The question that we can answer with our framework is: how much of SWC is due to the banking market (i.e., to the factors) and how much is due/driven to/by the single banks behaviors (i.e., by the idiosyncratics). Furthermore, by means of our spectral analysis in Section \ref{sec:spectral}, we can also decompose SWC (SWC due to common/idiosyncratics) according to the frequency response to shocks. 
\section{Spectral connectedness, approximating factor model and estimation strategy} \label{sec:spectral}
Inspired by \citet{barunik2018measuring}, we can extend the idea of Section \ref{sec_model} to the frequency domain. This is important in economics. In fact, shocks to the economic activity can affect variables at various frequencies, with various degrees of strength. Therefore, being able to disentangle further the financial connectedness ($C^H, C^{H}_{Mkt}, C^{H}_{Ids}$) into long/medium/short term response to shocks, appears of great practical relevance. The idea is rather simple: to describe the frequency dynamics of the connectedness, one can consider the spectral representation of variance decompositions based on \emph{frequency} responses to shocks, instead of impulse responses to shocks, as done thus far \citep[see also][Sec. 1.2]{barunik2018measuring}. As our proposed approximating model is an approximate static factor model, we have the following structure as the spectral analogue of \eqref{eq_decom}-\eqref{eq_irf}.  The population spectral density matrix, at frequency $\omega$, for the factor process is given by 
\begin{equation}\label{spectrum_factors}
\BS f_f(\omega)= \bigg[\sum_{h=0}^{\infty}\BS \Psi^{(h)}_f  \exp(-\mathrm{i} h \omega)\bigg] \BS \Sigma_u \bigg[\sum_{h=0}^\infty \BS \Psi^{(h)}_f \exp(\mathrm{i} h \omega)\bigg]^\top, \quad \omega \in [0,2\pi].    
\end{equation}\normalsize

Likewise, for the idiosyncratic component: 
\begin{equation}\label{spectrum_idios}
\BS f_{\xi}(\omega)= \bigg[\sum_{h=0}^{\infty}\BS \Psi^{(h)}_{\xi}  \exp(-\mathrm{i} h \omega)\bigg] \BS \Sigma_v \bigg[\sum_{h=0}^\infty \BS \Psi^{(h)}_{\xi} \exp(\mathrm{i} h \omega)\bigg]^\top, \quad \omega \in [0,2\pi].    
\end{equation}\normalsize
Here, the $\sum_{h=0}^\infty \BS \Psi^{(h)}_{f/\xi} \exp(-\mathrm{i} h \omega)$ are the Fourier transforms of the respective MA($\infty$) coefficients, where $\mathrm{i}=\sqrt{-1}$.  
Therefore, the spectral density (or ``power spectrum"), at frequency $\omega\in[0,2\pi]$, of the process $\{\BS x_t\}$ is given by
\small
\begin{equation}
\begin{aligned}
\label{eq.spectral.density.X}
    \BS f_x(\omega)&=\sum_{h=-\infty}^{\infty}\mathbb{E}(\bx_t\bx_{t-h}^{\top})\exp(-\mathrm{i}h\omega) \\
    &=\BS \Lambda \BS f_f(\omega) \BS \Lambda^\top +\BS f_\xi(\omega)\\ &=\BS \Lambda \bigg[\sum_{h=0}^{\infty}\BS \Psi^{(h)}_f  \exp(-\mathrm{i} h \omega)\bigg] \BS \Sigma_u \bigg[\sum_{h=0}^\infty \BS \Psi^{(h)}_f \exp(\mathrm{i} h \omega)\bigg]^\top \BS \Lambda^{\top}+\bigg[\sum_{h=0}^{\infty}\BS \Psi^{(h)}_{\xi}  \exp(-\mathrm{i} h \omega)\bigg] \BS \Sigma_v \bigg[\sum_{h=0}^\infty \BS \Psi^{(h)}_{\xi} \exp(\mathrm{i} h \omega)\bigg]^\top\\ &=\left[\sum_{h=0}^\infty\begin{pmatrix}
        \bLambda \BS \Psi^{(h)}_f & \BS \Psi^{(h)}_{\xi}
    \end{pmatrix}\exp(-\mathrm{i} h \omega)\right] \bSigma_{\eta} \left[\sum_{h=0}^\infty\begin{pmatrix}
        \BS \Psi^{(h) \top}_{f} \bLambda^{\top} &  \BS \Psi^{(h)\top}_{\xi}
    \end{pmatrix} \exp(\mathrm{i} h \omega)\right].
\end{aligned}
\end{equation}

The $\BS f_x(\omega)$ describes how the variance of $\bx_t$ is distributed over the frequency components $\omega$, where we note that $\mathbb{E}(\bx_t\bx_{t-h}^{\top})=\frac{1}{2\pi}\bigintssss_{-\pi}^{\pi}\BS f_x(\omega)\exp(\mathrm{i}h\omega) d\omega$. Interestingly, given our factor model decomposition of $\bx_t$, this variance distribution over the frequencies is disentangled into variance from the common component and variance from the idiosyncratic component. In fact, the \emph{generalized causation spectrum} over the frequencies $\omega\in(-\pi,\pi)$ can be defined as
\begin{equation}
    \begin{aligned}
    (\texttt{f}(\omega))_{kj}=\frac{\left(\BS e_j^{\top}\BS \Sigma_{\eta} \BS e_j\right)^{-1}\left|\sum_{h=0}^{\infty}\be_k^{\top}\left[\begin{pmatrix}
        \bLambda \BS \Psi^{(h)}_f &\BS \Psi^{(h)}_{\xi}
    \end{pmatrix}\exp(-\mathrm{i} h \omega) \right]\bSigma_{\eta}\be_j\right|^2}{\left(\be_k^{\top}\BS f_x(\omega)\be_k\right)}, 
    \end{aligned}
\end{equation}
\normalsize
for $k=1,\ldots,N, \; j=1,\ldots, r+N$ and where as before $\BS e_j (\BS e_k)$ is a selection vector with $j (k)$-th element unity and zeros elsewhere. Here, $(\texttt{f}(\omega))_{kj}$ is the spectral analogue of \eqref{theta}, and it measures the portion of the spectrum of the $k$th variable at frequency $\omega$, due to shocks in the $j$th variable. Now, in the same way as \citet{barunik2018measuring}, in order to obtain a decomposition of variance decompositions to frequencies, it is necessary to weight $(\texttt{f}(\omega))_{k,j}$ by the frequency share of the variance of the $k$th variable. Therefore, the weighting function can be defined as $(\Gamma(\omega))_k=\frac{\left( \be_k^{\top} \BS f_x(\omega)\be_k\right)}{\frac{1}{2\pi}\bigintsss_{-\pi}^{\pi}\left( \be_k^{\top} \BS f_x(\lambda)\be_k\right) d\lambda}$,
representing the power of the $k$th variable at a given frequency $\omega$, summing through frequencies to a constant value of $2\pi.$ It then finally follows that the spectral representation of the variance decomposition from $k$ to $j$ can be stated as 
\begin{equation}
    (\theta(\infty))_{kj}=\frac{1}{2\pi} \bigintssss_{-\pi}^{\pi}(\Gamma(\omega))_k(\texttt{f}(\omega))_{kj} d\omega.
\end{equation}\normalsize
Let us note that $\lim_{H\to \infty}\theta_{kj}^g(H)$ as in \eqref{theta} is a weighted average of the \emph{generalized causation spectrum} $(\texttt{f}(\omega))_{kj}$ which gives the strength of the relationship at frequency $\omega$, weighting by the power of the series on that frequency. Furthermore, to define connectedness at short/medium/long term frequencies, it is necessary to work with frequency bands. 
Hence, let us define a frequency band $d=(a,b):a,b\in(-\pi,\pi), a<b$, such that the FEVD on frequency band d ($\text{FEVD}_d$) can be defined as \begin{equation}\label{fgfevd}
    (\theta_d)_{kj}=\frac{1}{2\pi} \bigintssss_{d}(\Gamma(\omega))_k(\texttt{f}(\omega))_{kj} d\omega,\quad C_{k\leftarrow j}^{H,\omega}=\frac{(\theta_d)_{kj}}{\sum_j (\theta(\infty))_{kj}},
\end{equation} \normalsize
where $C_{k\leftarrow j}^{H,\omega}$ is its scaled version (to sum up to 1, as before).
With all this in place, we can explore the  
\emph{frequency} connectedness on a frequency band $d$, both for the factors, the idiosyncratics and the summation of both (system-wide).  
As in Section \ref{sec_model} for the GIRF and FEVD in \eqref{girf},\eqref{theta}, also in the case of the $\text{FEVD}_d$ in \eqref{fgfevd} the distinction between market and idiosyncratics connectedness is obtained by summing $(\texttt{f}(\omega))_{kj}$  over $j=1,\ldots,r$ for ``market only", and $j=r+1,\ldots, r+N$ for ``idiosyncratics only".

Table \ref{tab:spectral_conn} below summarizes these measures:

\begin{table}[h!]
\centering
\begin{threeparttable}
\begin{tabular}{c|l|}

 &  \textbf{Frequency Connectedness} 
 \\ [2ex]
 
\textbf{Market} & $C_{{Mkt},d}^{H,\omega}=N^{-1}\sum_{k=1}^N\sum_{\substack{j=1}}^r C_{k\leftarrow j}^{H,\omega}\frac{\sum \theta_d}{\sum \theta(\infty)}$ \\[2ex]

\textbf{Idiosyncratics} &$C_{{Ids},d}^{H,\omega}=N^{-1}\sum_{k=1}^N\sum_{\substack{j=r+1\\j-r\neq k}}^{r+N} C_{k\leftarrow j}^{H,\omega}\frac{\sum \theta_d}{\sum \theta(\infty)}$\\[2.5ex]

\textbf{System-wide} 
&$C_{d}^{H,\omega}=C_{{Mkt},d}^{H,\omega}+C_{{Ids},d}^{H,\omega}$ \\[2ex]
\hline
\end{tabular}
\caption{Spectral Connectedness}
\label{tab:spectral_conn}
\end{threeparttable}
\end{table}
where $\sum \theta_d, \sum \theta(\infty)$ stands for the sum of all elements of $\theta_d, \theta(\infty)$, respectively.
\subsection{Spectral Estimation}
In order to obtain an in-sample estimate of \eqref{eq.spectral.density.X}, both spectral densities for the factors, $\bof_f(\omega)$, and for the idiosyncratics, $\bof_{\xi}(\omega)$, need to be estimated.
 
As $r$ is of fixed dimension, the spectral density $\BS f_f(\omega)$ (or its inverse) can be estimated by classical methods such as non-parametric lag-window estimators \citep[see a.o.,][and the specifics in Section \ref{sec_techdet}]{wu2018asymptotic}. For the idiosyncratics, we can explicitly use the VAR structure to estimate their spectral density. We note though how the natural estimator is the \emph{inverse} spectral density and an additional assumption of \emph{column-wise} sparsity of the VAR companion-form matrix is required (see Section \ref{sec_techdet}, Assumption \ref{sparsity.b}). This is an additional requirement when estimating spectral densities, in fact in Section \ref{sec_model}, as in \citet{krampe2021factor}, we only required \emph{row-wise} sparsity (see Section \ref{sec_techdet}, Assumption \ref{sparsity}). Additionally, a parametric estimation of the spectral density matrix of a VAR process requires an estimate of the covariance, or precision matrix, of the residual process $\{\BS v_t\}$, i.e., $\BS \Sigma_v$ or $\BS \Sigma_v^{-1}$ in \eqref{spectrum_idios}.
The residuals can be consistently estimated by $\hat {\BS v}_t=\hat {\BS \xi}_t-\sum_{j=1}^{p_{\xi}} \hat {\BS B}^{(j)} \hat {\BS \xi}_{t-j}, t=p_{\xi}+1,\dots,T$. Then, based on these, procedures such as the graphical LASSO of \cite{friedman2008sparse}, (A)CLIME of \cite{cai2011constrained} or fused LASSO of \citet{dallakyan2023fused} can be used in order to obtain a regularized estimator of $\bSigma_v$ or $\bSigma_v^{-1}$ (respectively, we will employ the notation $\hat{\bSigma}_v^{(re)}$, $\hat{\bSigma}_v^{(re)-1}$, where ``re" is a shorthand for ``regularized").  
With this, we get the following estimator for $\BS f_\xi(\omega)^{-1}$: 
\begin{align}\label{spect_idio}
    \hat{\BS f}_\xi(\omega)^{-1}=\bigg[\BS I_N-\sum_{h=1}^{p_{\xi}} \hat{\BS B}^{(thr,h)} \exp(\mathrm{i} h \omega)\bigg] \hat{\bSigma}_v^{(re){-1}} \bigg[\BS I_N-\sum_{h=1}^{p_{\xi}} \hat{\bB}^{(thr,h)} \exp(-\mathrm{i} h \omega)\bigg]^\top,
\end{align}\normalsize
where $\hat{\BS B}^{(thr,h)}=\THRarg{\lambda_\xi}(\hat{\BS B}^{(h)})$ and $\THRarg{\lambda_\xi}(\cdot)$ is a thresholding function with threshold parameter $\lambda_\xi$,  fulfilling the conditions $(i)$ to $(iii)$ in Section 2 in \citet{cai2011adaptive}. For instance, such a thresholding  function can be the adaptive LASSO thresholding function given by $\THRarg{\lambda_\xi}^{al}(z)=z(1-|{\lambda}/z|^\nu)_+$ with $\nu\geq1$. Soft thresholding ($\nu=1$) and hard thresholding ($\nu=\infty$) are boundary cases of this function. These thresholding functions act by thresholding every element of the matrix $\hat{\BS B}^{(h)}$ resulting in a row- and column-wise consistent estimation of the VAR slope matrices. In Section \ref{sec_techdet}, Lemma~\ref{lem.rate.spectral.var}, we present the error bounds for $\|\hat{\BS f}_\xi(\omega)^{-1}-{\BS f}_\xi(\omega)^{-1}\|_\infty$ and $\|\hat{\BS f}_\xi(\omega)^{-1}-{\BS f}_\xi(\omega)^{-1}\|_2$. Inversion of \eqref{spect_idio} yields the estimator $\hat{\BS f}_\xi(\omega)$. Finally, replacing in $\bof_x(\omega)=\BS \Lambda \bof_{f}(\omega)\BS \Lambda^{\top}+\bof_{\xi}$ both estimated spectral densities discussed above leads to our final estimator of the spectral density matrix $\hat{\bof}_x(\omega)$. Its error bounds $\|\BS f_x(\omega)-\hat {\BS f}_x(\omega)\|_l$ for $l=1,2,\infty$ are presented in Section \ref{sec_techdet}, Theorem~\ref{thm.spectral.density}.

\section{Data \& Results}\label{sec_data}
We make use of two datasets comprising a large number of global bank assets.  \begin{itemize}
    \item[(i)] First, we employ the dataset provided in \citet{demirer2018estimating} containing stock price volatilities for 96 banks from 29 developed and emerging economies, plus the bond price volatilites of 10 major world countries.\footnote{The countries are: USA, UK, Germany, France, Italy, Spain, Greece, Japan, Canada, Australia.} This is daily data spanning from September 12, 2003 until January 30, 2014. 
    \item[(ii)] Second, we compute and employ a more recent vintage of (i), spanning daily from February 20, 2014, up until June 14, 2023.
\end{itemize}
 Dataset (ii) necessarily has some differences with respect to (i). In fact, it comprises 83 stock price volatilities (instead of 96). The remaining ten series are the bond price volatilities of the same ten major world countries as in (i). The reason for the lack of 13 banks in (ii) with respect to (i) is that certain banks considered before are either not traded anymore in the new sample or they have too many missing values. We provide a complete list in Table \ref{Bank List}. Stock prices are from Datastream and Bond prices are from Bloomberg. To compute daily range-based realized volatilities\footnote{This type of volatility is the same computed in \citet{demirer2018estimating} and is almost as efficient as realized volatility based on high-frequency intra-day data given it is robust to certain forms of micro structure noise, see \citet{alizadeh2002range}.} we use the formula below, namely  \begin{align}
    \sigma^2_{i,t}&=0.511(H_{i,t}-L_{i,t})^2-0.019 [(C_{i,t}-O_{i,t})(H_{i,t}+L_{i,t}-2O_{i,t})\\&-2(H_{i,t}-O_{i,t})(L_{i,t}-O_{i,t})]-0.383(C_{i,t}-O_{i,t})^2,\notag
\end{align}\normalsize
where $H_{i,t}, L_{i,t}, O_{i,t}, C_{i,t}$ are the logs of daily high, low, opening and closing prices for bank stock $i$ on day $t$. We are going to focus on \emph{system-wide connectedness} $C^H$, {for $H=10$}, and compute the part of it due to the (banking) market: $C^H_{Mkt}$, and the part due to the idiosyncratic shocks $C^H_{Ids}$, such that $C^H=C^H_{Mkt}+C^H_{Ids}$. 
Following \citet{demirer2018estimating}, we employ a rolling window of 150 days and the reporting time point corresponds to the final day of the window. We estimate the factors and loadings via PCA, selecting the number of factors and lag-length of the VAR using the extended BIC information criteria of \citet{krampe2021factor} (see also Section \ref{sec_techdet}). This gives us for both Dataset (i) and Dataset (ii) only one common factor ($r=1)$ and $p_f=2$. The idiosyncratics are estimated via adaptive LASSO where initial weights are preliminary plain LASSO weights and the lag-length is estimated to be $p_{\xi}=4$.\footnote{In Section \ref{sec_techdet} we show that $r,p_f,p_{\xi}$ can be jointly obtained via minimization of a single information criterion as in \citet{krampe2021factor}.} The LASSO tuning parameter $\lambda$ is selected via standard BIC (see \citealp{hecq2023granger} for an overview of data-driven techniques to select the tuning parameter). Together with the SWC measure, we also report $95\%$ Bootstrap confidence bands. The full details of their computation are given in Section \ref{sec_boot}.
\subsection{\textbf{Dataset (i): 2003-2014}}
\begin{figure}[htbp]
    \centering
\includegraphics[page=2, width=0.95\textwidth, height=7cm]{Result_up_to_2013.pdf}
\includegraphics[page=1, width=0.95\textwidth, height=7cm]{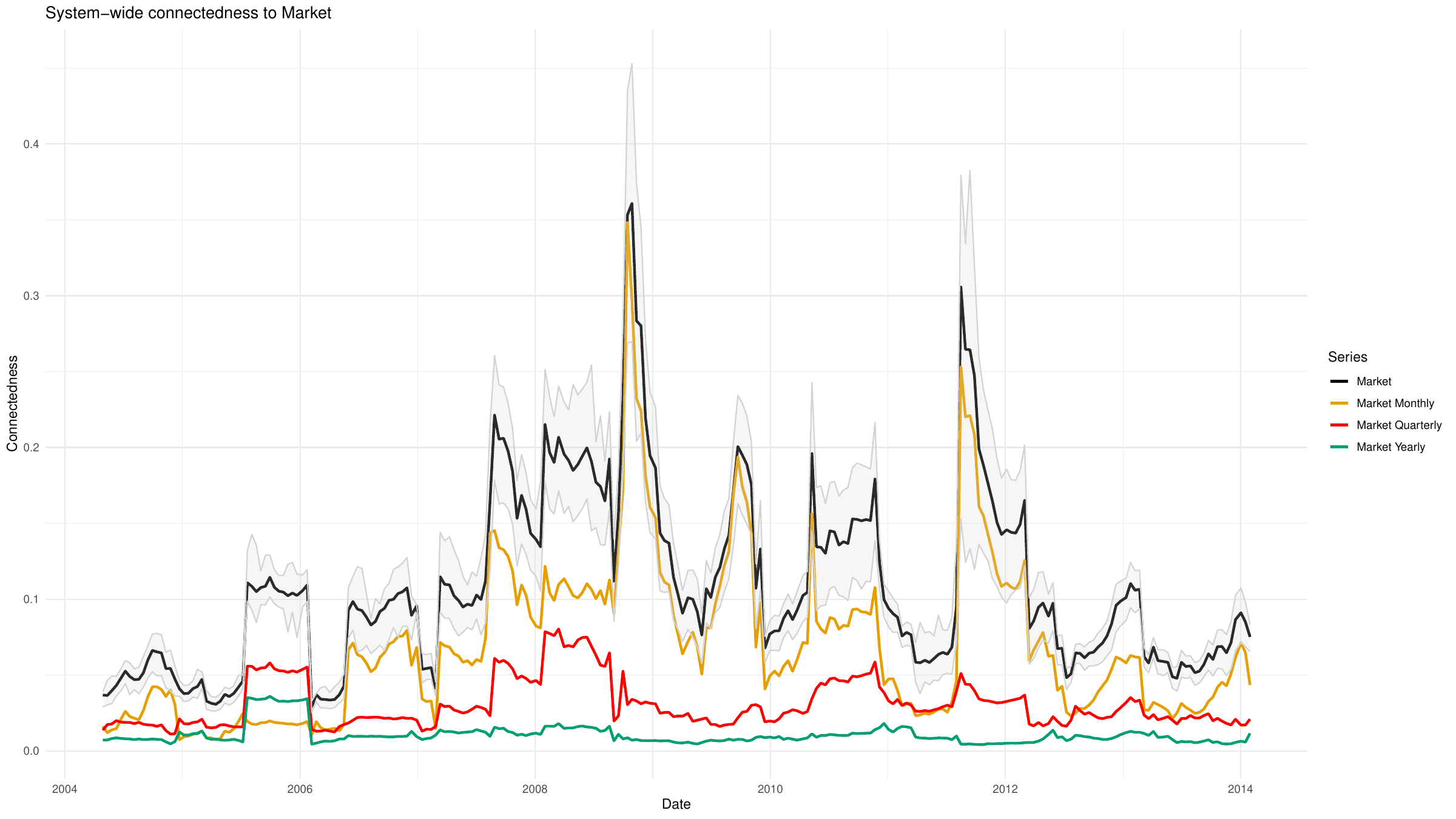} 
\includegraphics[page=3, width=0.95\textwidth, height=7cm]{Result_up_to_2013.pdf}\caption{Top panel: System-Wide Connectedness ($C^H$); Center panel: System-Wide Connectedness due to Market ($C^H_{Mkt}$); Bottom panel: System-Wide Connectedness due to Idiosyncratics ($C^H_{Ids}$). Frequency band: Monthly (orange), Quarterly (red), Yearly (green); Confidence bands (grey shade). Span: 2003-2013, 150 days rolling window.}\label{figure_0314}
\end{figure}

Starting from $C^H$ in Figure~\ref{figure_0314}, 
we find an entirely similar shape, and roughly the same magnitude, as the SWC computed in \citet{demirer2018estimating} (see especially their Figure 9). With the Federal Reserve decision to tighten monetary policy in May-June 2006, we observe an upward trending behavior of the SWC, culminating in the Lehman bankrupcy in 2008. Indeed, Lehman Brothers filed for bankruptcy on September 15, 2008. At that time, our estimated SWC is found on the rise, where on September 16th reaches a level of 74.7\%, only to continue towards its highest peak reached on November 25th, at a staggering 89.7\% SWC. It will take the whole year of 2009 for the SWC to reassess at a pre-crisis level of roughly 70\% SWC. Two other notable SWC jumps correspond to May 2010, due to delays in Greece's rescue package, and another in August 2011, as sovereign debt and banking sector concerns spread to Spain and Italy.
While the magnitude of SWC is in the same ballpark as \citet{demirer2018estimating}, it is though slightly (roughly 5\%) higher, and this is especially visible in calmer times.\footnote{Take the beginning of the sample for instance, our estimated SWC starts at a level of 64.8\%, while \citet{demirer2018estimating} estimates it just under the 60\% threshold.} One likely reason of this is that the VAR Elastic Net of  \citet{demirer2018estimating} directly ``sparsifies" the number of banks in every estimate of the VAR equations, while we only shrink part of the connections among different cross-sections. This entails that if truly factors are playing a role, then such direct sparsity is also implicitly shrinking the loadings. In our case, the factor model does \emph{not} impose direct sparsity on the linkages of $\bx_t$, nor on the loadings, but retains one strong factor representing the common behavior of \emph{all} banks. Instead, we just sparsify the idiosyncratics' dynamic which, as discussed, is a more reasonable consideration. In other words, the assumption of sparsity of \citet{demirer2018estimating} can potentially lead to underestimation of the degree of connectedness if truly the data linkages are many, and if the factors are strong. Furthermore, the fact that SWC is generally 'quite high' in our results, resonates well with the findings in a.o., \citet{allen2000financial, acemoglu2015systemic}. Namely, that when the magnitude of the shocks is below a certain threshold, ``a more diversified pattern of inter bank liabilities leads to a less fragile financial system". The other way around, if the shocks' magnitude surpasses a certain threshold ``highly diversified lending patterns facilitate financial contagion and create a more fragile system" \citep[][pg. 566]{acemoglu2015systemic}. These last considerations provide an interesting connection with the spectral analysis. In fact, further decomposing SWC into frequencies, thus considering our spectral-SWC $C_{d}^{H,\omega}$ for $\omega=\{\text{Monthly, Quarterly, Yearly}\}$, it so appears that a medium frequency response to shocks, $\omega=\{\text{Quarterly}\}$, seems to dominate the calmer times, while the more rapid monthly frequency, $\omega=\{\text{Monthly}\}$, leads and overcomes during the crises. The spectral SWC at monthly frequency indeed reaches its highest peak on October 28th 2008 (slightly leading the main peak of the SWC), where it reaches a value of 0.465, i.e., 52.3\% of the SWC at the same date (0.889). The remaining 48\% is made up of the other frequencies, respectively quarterly (0.271) and yearly (0.162).
Equally interesting is to observe by what types of shocks (market/idiosyncratics), when, and at which frequency, the SWC is driven, i.e., looking at $C^H_{Mkt}, C^H_{Ids}$, and the respective spectral versions $C^{H,\omega}_{Mkt,d}, C^{H,\omega}_{Ids,d}$. Interestingly, we observe how the connectedness is \emph{mostly} driven by idiosyncratic variation. In fact, $C^H_{Ids}$ averages at 0.6, meaning on average drives 80\% or more of the SWC, while only 20\% or less is left to the common component $C^H_{Mkt}$. However, $C^H_{Mkt}$ does jump upward during crises, or more generally financially turbulent times. We can observe how during the 2008 crisis, on October 28th, the $C^H_{Mkt}$ reaches its maximum peak at 0.36, i.e., roughly 40.5\% of the whole SWC at the same date (0.889). While it doubles if compared to pre crisis levels, it is still not driving the majority of the SWC, which is indeed driven 59.5\% by $C^H_{Ids}$. In terms of frequency response to shocks, we observe how $C^H_{Mkt}$ is driven for the major part by short frequencies, i.e., by $C^{H,\omega}_{Mkt,d}$, for  $\omega=\{\text{Monthly}\}$, whilst the longer frequencies are almost irrelevant (especially, the yearly one). Interestingly, the opposite shall be said about the frequencies response decomposition for $C^H_{Ids}$. Indeed, we find that $C^H_{Ids}$ is predominantly driven by the longest frequency $C^{H,\omega}_{Ids,d}$,  $\omega=\{\text{Yearly}\}$, which accounts alone for about 50\% of $C^H_{Ids}$, whilst the quarterly and yearly frequencies accounts for less than 30\% and 20\%, respectively.
Overall, we see that, given only one estimated common factor, SWC is driven predominantly, and in the long-run, by idiosyncratic connectedness. Crisis times instead see a surge in common component connectedness, driven purely by short-run dynamics.

\subsection{\textbf{Dataset (ii): 2014-2023}}
\begin{figure}[htbp]
    \centering
    \includegraphics[page=2, width=0.95\textwidth, height=7cm]{Result_from_2014.pdf}
\includegraphics[page=1, width=0.95\textwidth, height=7cm]{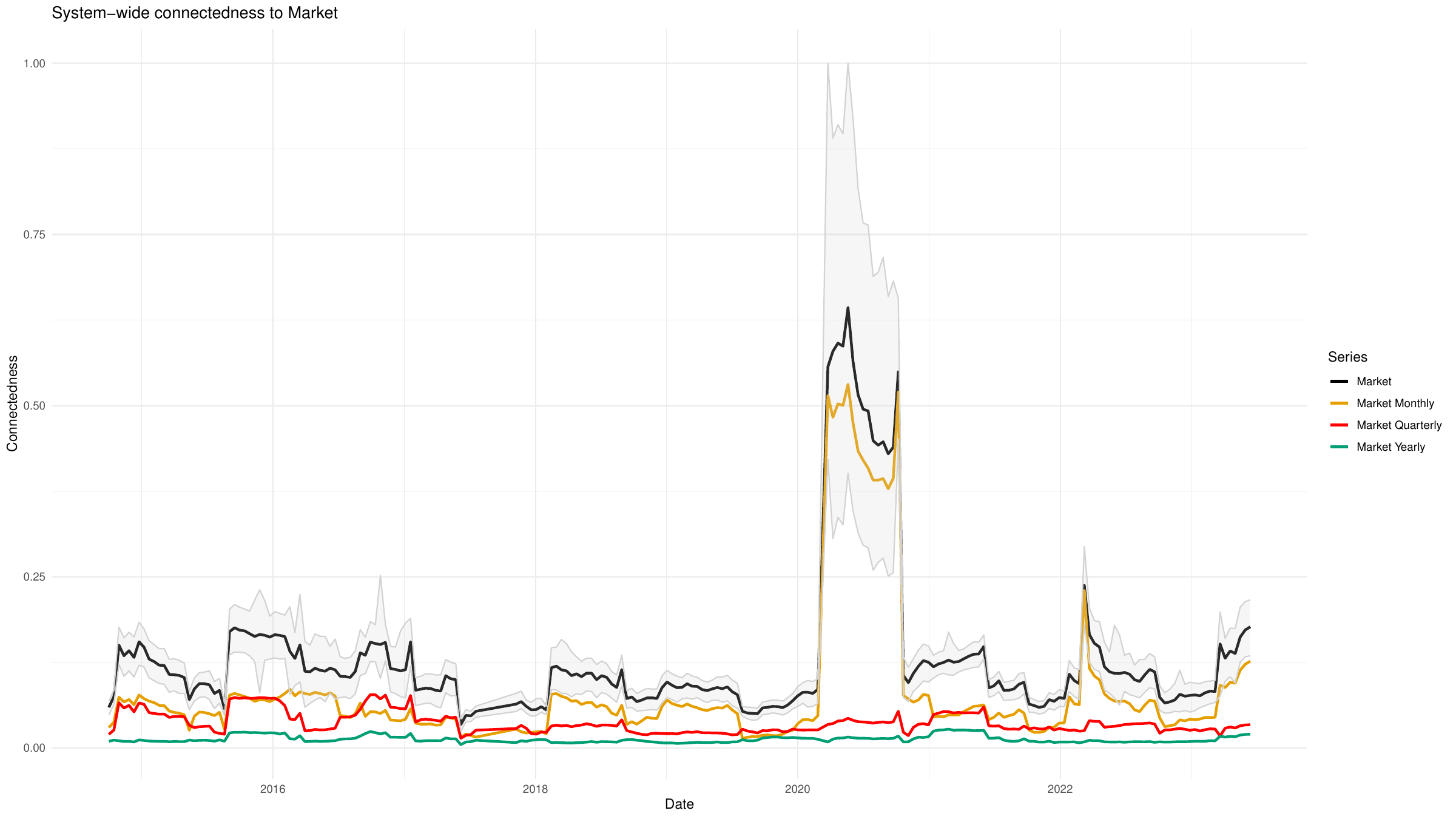}
\includegraphics[page=3, width=0.95\textwidth, height=7cm]{Result_from_2014.pdf}\caption{Top panel: System-Wide Connectedness ($C^H$); Center panel: System-Wide Connectedness due to Market ($C^H_{Mkt}$); Bottom panel: System-Wide Connectedness due to Idiosyncratics ($C^H_{Ids}$). Frequency band: Monthly (orange), Quarterly (red), Yearly (green); Confidence bands (grey shade). Span: 2014-2023, 150 days rolling window.}\label{fig_1423}
\end{figure}
Now we discuss the connectedness results in Figure \ref{fig_1423} on our more recent (2014-2023) dataset, containing almost the same variables (some are discontinued, see Table \ref{Bank List} for a list) as in the previous analysis. The level of SWC (0.69) picks up from where was left after January 2014 in the previous analysis (0.67). $C^H$ exhibits less of a clear trending behavior in this new sample period but more of a level-shift/oscillatory one, around 75\%. This level is roughly 10\% more than the pre-2008 crisis, as it seems that the lesson of 2008 resulted in a more cohesive inter-bank network. A remarkable shift of the SWC is observed already between 2015 and 2017, where the mean wanders around 80\% with heights of 84\%. Several events can be associated with this raise of the SWC: from the European banking sector problems (a.o., Greece's debt crisis, italian banks' loans crisis), to low interest rates and negative rates in Europe, Brexit, the Chinese economic slowdown and stock market crash of 2015, just to name a few. 
The main, hard-to-miss event of relevance within the sample is of course the Covid19 outbreak and the consequent global crisis starting in 2019. In fact, we observe how in correspondence of the Covid19 outbreak, the SWC exhibits a vertical increase of more than 20\%, from 72\% (February 2020) to over 95\% (March 2020). This unprecedented SWC level is maintained roughly until October 2020, after which an equally vertical drop is observed back to a level of roughly 75\% or less. While a relapse that touches 80\% can be observed in March 2022, in the remaining part of our sample SWC never reaches similar heights as March-October 2020. By observing the decomposition of SWC into frequencies, an entirely analogous picture as for the previous analysis presents itself. The medium frequency response to shocks, $\omega=\{\text{Quarterly}\}$, dominates the calmer times, even during 2015-2017, while the more rapid monthly frequency, $\omega=\{\text{Monthly}\}$, has a vertical increase of almost 50\%, from 0.17 on 26th October 2020, to 0.60 in May and even 0.61 in July of the same year, only to drop back to a level of 24\% and less from October 2020 onwards, with a short relapse at around 37\% in March 2022. Disentangling again SWC into $C^H_{Mkt}$ and $C^H_{Ind}$, we again see how SWC is mostly driven by idiosyncratic variation. In this second sample, $C^H_{Ids}$ averages at roughly 0.65 (5\% more compared to the previous sample) corresponding to more than 85\% of the whole SWC, leaving only the remaining 15\% to $C^H_{Mkt}$. The latter is again predominantly driven by the short frequency response to shocks ($C^{H,\omega}_{Mkt,d}$, for  $\omega=\{\text{Monthly}\}$) and spikes upward during crisis. Especially, in the case of the abnormal circumstances of Covid19 we can observe almost an overlap between $C^H_{Mkt}$ and $C^{H,\omega}_{Mkt,d}$, for  $\omega=\{\text{Monthly}\}$. Vice-versa, $C^H_{Ids}$ is again mostly driven by the longest frequency (yearly), but one difference can be observed during the Covid19 outbreak. While for the 2008 crisis the idiosyncratic connectedness did not drop dramatically, it is most definitely the case for the Covid19. While the confidence bands are wide, in May 20th, 2020, $C^H_{Ids}$ is estimated at 0.31 (only 32\% of the SWC) while $C^H_{Mkt}$ is at 0.64 (67\% of SWC), thus practically inverting the levels of one-another.
\subsection{Comparison of Major Crisis}
In the previous analyses we observed two, very different global crises: the 2008 subprime crisis and the 2019 Covid19 pandemic. The former is a financial collapse rooted in the financial sector and its dynamics; the latter is a global health crisis that had unprecedented effects (for the recent history, at least) on the economy and the financial stability of countries and institutions. Both of these, in their own ways, are responsible for an increased level of global economic uncertainty, and in some cases of proper panic-spreading in the financial markets (aside of elsewhere). We have seen how, in line with \citet{demirer2018estimating}, global crises correspond to an increase in the overall SWC. However, we have also been able to uncover how, when times are calm, SWC is predominantly driven by an idiosyncratic dynamic. When global crises hit instead, a sharp increase in the connectedness due to the common component (market dynamic) is observed, in line with e.g., \citet{barigozzi2021time}, and this is almost entirely driven by a short dynamic response to shocks. While both crises are in their own way somewhat unprecedented events, the financial crisis in 2008 has undoubtedly quite a different shape with respect to the Covid19 pandemic. Banks seem to have perceived the 2008 collapse quite some years in advance, as visible from the building-up pattern of $C^H$ from late 2004 onwards, all the way to 2008 \citep[see also][]{barrell2008evolution}. It also seems that such a crisis has had a more ``sticky" effect for the connectedness, which ever since has maintained a slightly higher level of interconnections than before. The Covid19 crisis instead, has created an unprecedented vertical increase in SWC and SWC due to the Market, and correspondingly the most vertical drop in $C^H_{Ids}$. The global panic generated by such a crisis has very rapidly shoot up $C^H$. This finding is in line with e.g., \citet{bouri2021return}, who also find Covid19 has altered the network of (in their case return) connectedness by generating sudden increases in the system-wide connectedness. The reached peak of $C^H$ is then maintained roughly for the whole duration of the uncertainty created by the pandemic and the connectedness then recovers --almost at the same vertical pace-- the pre-crisis level.

\begin{remark}\label{rmk_1}
For comparison purposes with \citet{demirer2018estimating}, we chose a rolling window of 150 days for estimating connectedness. This choice of a not-too-large window is also dictated by the need of ensuring stationarity of the volatilities, since we do not assume local stationarity. It follows that one should interpret with some care the big vertical surges/drops relative to the time stamp, since this might be due to the rolling window not covering the regime shift. Furthermore, regarding the frequency domain analysis, the choice of this window does affect the interpretation of what one would perhaps call ``long run" response to shocks. After all, 150 days is only half a year, hence this would not be sufficient to capture business-cycle like fluctuations.
\end{remark}

\begin{remark}\label{rmk_2}
The data employed, paired with the criterion we adopted to select the number of factors, has returned only one estimated factor. Now, while the data driven estimation procedure is robust, according to the finite sample results in \citet{krampe2021factor}, one might wonder if an additional factor would change the empirical analysis substantially. For instance, some might argue that a factor representing the exchange rate between the various currencies of the stock prices might be sensible to add. For this, we replicate the analysis with 2 factors, whose results are reported in Figure \ref{figure_app1} in the Appendix. By virtue of the factor model decomposition, it is clear that adding more factors will increase the share of total connectedness due to the common component, in turn reducing the one due to the idiosyncratics. With two factors, we observe a 5-10\% increase in connectedness due to common components, and a corresponding 5-10\% decrease in idiosyncratic connectedness. However, the main conclusions remain unchanged: during stable periods, total connectedness is primarily driven by idiosyncratic factors, while in crises, connectedness due to the common component increases and may dominate.
\end{remark}
\section{Technical Details}\label{sec_techdet}
A few words on the notation we employ. For any vector $\boldsymbol{x}\in \mathbb{R}^n$, $\norm{\boldsymbol{x}}_p = \left(\sum_{i=1}^n |x_i|^p \right)^{1/p}$ denotes the $\ell_p$-norm and $\be_j$ denotes a unit vector of appropriate dimension with the one in the $j$th position. For a $r\times s$ matrix $\BS A=(a_{i,j})_{i=1,\ldots,r, j=1,\ldots,s}$,  $\|\BS A\|_1=\max_{1\leq j\leq s}\sum_{i=1}^r|a_{i,j}|=\max_j \| \BS A \be_j\|_1$, $\|\bA\|_\infty=\max_{1\leq i\leq r}\sum_{j=1}^s|a_{i,j}|=\max_{i} \| \be_i^\top \BS A\|_1$ and $\|\BS A\|_{\max}=\max_{i,j}\allowbreak |\be_i^\top \BS A \be_j|$.
$\bA^i$ denotes the $i$th matrix power of $\bA$ and $\bA^{(i)}$ refers to the $i$th element of a sequence of matrices. We denote the largest/smallest absolute eigenvalue of a square matrix $\BS A$ by $\sigma_{\max/\min}(\BS A)$ and $\|\BS A \|_2^2=\sigma_{\max}(\BS A \BS A^\top)$.
$\|\BS x\|_0$ denotes the number of non-zero elements of $\BS x$.  $\plim$ denotes convergence in probability.
\subsection{Details on Estimation Procedure - Time Domain}\label{sec_detest} In estimating the dynamic factor model with sparse VAR idiosyncratic components we closely follow the work of \citet{krampe2021factor}. We report here a summary of the main assumptions and, importantly, the estimation algorithm. We refer to the said paper for details.
\par We work with factors $\{\bof_t\}$ and idiosyncratics $\{\bxi_t\}$ being second order, uncorrelated stationary processes, both with bounded $\ell_2$ innovation covariances, and with the idiosyncratics autocovariance matrix bounded in $\ell_2$ norm for increasing $N$. Eight finite moments are assumed on the innovation process $\{(\bu_t^{\top},\bv_t^{\top})^{\top}, t\in\mathbb{Z}\}$ and weak factors are ruled out, so each of the factors provides a non-negligible contribution to the variance of each component of $\{\bx_t\}.$ For the idiosyncratics VAR($p_{\xi}$) coefficient matrix, approximate {row-wise} sparsity is assumed and it is allowed to grow with the sample size. The following Assumption \ref{sparsity}-Assumption \ref{ass.fac} formalize this, where we use the notation $M_1,\ldots,M_8$ to refer to some positive constants.

\begin{assumption}\label{sparsity}(\textit{Sparsity and stability})\\
$(i)$ Let $\A$  denote the stacked (companion) VAR matrix of $\bxi_t$. Let $k$ denote the row-wise sparsity of $\A$ with approximate sparsity parameter $q \in [0,1)$, i.e.,\footnote{{$q=0$ corresponds to the \textit{exact sparsity} assumption where several parameters are exactly zero. \textit{Approximate sparsity} $q>0$ allows for many parameters not to be exactly zero but rather small in magnitude.}} 
$$\max_i \sum_{s=1}^{p_{\xi}} \sum_{j=1}^N |\BS B_{i,j}^{(s)}|^q=\max_i \sum_{j=1}^{Np} |\A_{i,j} |^q\leq k.$$\normalsize
$(ii)$ The VAR process is considered as stable such that for a constant $\rho \in (0,1)$ we have independently of the sample size $T$ and dimension $N$: $\|\A^j\|_{2}=\sqrt{\sigma_{\max}(\A^{j\top}\A^{j})}\leq M_1 \rho^j$.\\ 
$(iii)$ We have $\|\BS \Gamma_\xi(0)\|_\infty\leq k_\xi M_2$, where $\BS \Gamma_\xi(0)=\var(\BS \xi_t)$ and $\sigma_{\min}(\var((\BS \xi_t^\top,\dots,\BS\xi_{t-p+1}^\top)^\top))>\alpha>0$.\\
$(iv)$ The parameters $k$, $k_\xi$ in $(i)$,$(iii)$ are allowed to grow with the sample size. 
\end{assumption}

\begin{assumption}\label{ass.moments} (\textit{Factor dynamics and  innovation moments of $\BS f_t, \BS \xi_t$})\\
$(i)$ The factors are given by a linear process with geometrically decaying coefficients: \begin{equation*} \boldsymbol{f}_t=\sum_{j=0}^\infty \BS D^{(j)} \bu_{t-j}, \end{equation*}\normalsize where $\|\BS D^{(j)}\|_2\leq K \rho^j$, for some constant $K>0$ and $\rho\in(0,1)$.\\
$(ii)$ $\{(\bu_t^\top,\bv_t^\top)^\top, t \in \Z\}$ is an i.i.d.~sequence with $\zeta> 8$ finite moments, i.e.,  $\mathbb{E} |\bu_{t,j}|^\zeta\leq M_3$ and
$ \max_{\|\bw\|_2\leq 1} \mathbb{E}  |\bw^\top \boldsymbol{v}_t|^\zeta\leq M_4$. Also, $\cov(\bu_t,\bv_t)=0$. 
\end{assumption}

\begin{assumption} \label{ass.fac}(\textit{Factors and loadings})\\
 $(i)$ $\plim_{T\to \infty} 1/T\sum_{t=1}^T \boldsymbol{f}_t \boldsymbol{f}_t^\top\!=\mathbb{E}( \boldsymbol{f}_t \boldsymbol{f}_t^\top)=\bSigma_F \in \R^{r\times r}$ positive definite, $\|\bSigma_F\|_2\leq M_5$. \\
$(ii)$  $\lim_{N \to \infty} 1/N \sum_{i=1}^N \bLambda_i \bLambda_i^\top=\bSigma_\Lambda \in \R^{r\times r}$, positive definite with $\sigma_{\Lambda,\max}\leq M_6$ and $\sigma_{\Lambda,\min}\geq 1/M_7>0$, $\|1/N \sum_{i=1}^N \bLambda_i \bLambda_i^\top\|_2\leq M_8$ for all $N$.\\
$(iii)$  All eigenvalues of $\bSigma_F,\bSigma_\Lambda$ are distinct.
\end{assumption}

\par For the estimation algorithm it is convenient to stack $\boldsymbol{x}_t,$ $t=1,\dots,T$ row-wise in order to obtain $\boldsymbol{X}=\boldsymbol{\chi}+\boldsymbol{\Xi}$ as a $T\times N$ matrix form of the factors \& idiosyncratics decomposition $\boldsymbol{x}_t=\bchi_t+\bxi_t$. The two step estimation procedure then proceeds as follows: \begin{enumerate}
    \item[(1.)] Perform a singular value decomposition of \begin{align}
    \label{eq.svd}\boldsymbol{X}/\sqrt{NT}=\BS U_{NT,r} \BS D_{NT,r} \BS V_{NT,r}^\top+\BS U_{NT,N-r} \BS D_{NT,N-r} \BS V_{NT,N-r}^\top,\end{align} \normalsize where $\BS D_{NT}$ is a diagonal matrix with the singular values arranged in descending order on its diagonal, $\BS U_{NT}$ and $\BS V_{NT}$ are the corresponding left and right singular vectors.  $\BS U_{NT,r} \BS D_{NT,r} \BS V_{NT,r}^\top$ corresponds to the $r$ largest elements in $\BS D_{NT}$, $\BS U_{NT}$ and $\BS V_{NT}$.  \bigbreak
     Set $\hat{\boldsymbol{F}}=\sqrt{T} \bU_{NT,r}$, $\hat{\boldsymbol{\Lambda}}=\sqrt{N} \BS V_{NT,r} \BS D_{NT,r}$, and $\hat {\BS \xi}=\BS X-\hat{\boldsymbol{F} }\hat{\boldsymbol{\Lambda}}^\top $.\\ The VAR($p_f$) for the factors is then given by: $\hat \bof_t=\hat{\boldsymbol{\Lambda}}\sum_{j=1}^{p_f}\BS D^{(j)} \hat{\BS f}_{t-j}+\bu_t.$
    \item[(2.)] Let $\hat {\BS{\xi}}_t^v=(\hat {\BS\xi}_t^\top,\dots,\hat {\BS\xi}_{t-p_{\xi}}^\top)^\top$. Then, an adaptive LASSO estimator for $\bbeta^{(j)}$ i.e., the $j$th row of $(\bB^{(1)},\dots,\allowbreak \bB^{(p_{\xi})}),$ is given by 
\begin{align}
    \label{eq.LASSO.beta.j}
    \hat{\bbeta}^{(j)}=\argmin_{\bbeta \in \R^{Np_{\xi}}} \frac{1}{T-p_{\xi}} \sum_{t=p_{\xi}+1}^T (\hat \xi_{j,t}-\bbeta^\top \hat {\BS\xi}_{t-1}^v)^2+\lambda \sum_{i=1}^{Np_{\xi}} |g_i \beta_i|,\quad j=1,\dots,N, 
\end{align}\normalsize
            where $\lambda$ is a non-negative tuning parameter which determines the strength of the penalty and $g_i, i=1,\dots,Np_{\xi},$ are weights.\footnote{{One can set $g_{i}=|\dot \beta_{i}|^{-\tau}$, where $\tau>0$ and $\dot \beta_{i}$ is an initial coefficient estimate. This is the \textit{adaptive} part of the LASSO problem. By setting the weights in this particular way, coefficients with high initial estimates receive proportionally low penalties. OLS can be used to obtain $\dot \beta_{i}$ but only if $Np_{\xi}<T$.}} For instance, $g_i=1$ leads to the standard LASSO. Let also $(\hat \bB^{(1)},\dots,\hat \bB^{(p_{\xi})})$ be the matrices that correspond to stacking $\hat {\bbeta}^{(j)},j=1,\dots,N$. To select $\lambda$ standard Bayesian information criterion is used.
    \item[(3.)] In order to select $r, p_f$ in (1.) and $p_{\xi}$ in (2.) the following joint extended information criterion is minimized. Let $Pen=\big(rp_f+\sum_{j=1}^{p_{\xi}} \|\be_i^\top \hat{\BS B}^{(j)}\|_0\big)\frac{\log(T)}{T} C_T$, for $i=1,\ldots,N$ and $C_T=c\frac{\log(NT/(N+T))}{\log(T)}$ with $c=1/2$, then 
\begin{align}\label{crit_FactLags}
    IC_{T,N}^{(i)}:=&\underset{r,p_{\xi},p_f}{\argmin}\;\ln\bigg[ {\frac{1}{T}\!\!\!\!\sum_{t=1+\max(p_{\xi},p_f)}^T \!\!\bigg(x_{i,t}
    -\sum_{j=1}^{p_f} \hat{\BS \Lambda}_i^\top \hat{\BS D}^{(j)} \hat{\BS f}_{t-j}
    -\sum_{j=1}^{p_{\xi}} \be_i^\top \hat{\bB}^{(j)}\hat{\bxi}_t^{(r)}\bigg)^2}\bigg]+Pen.
\end{align}
\end{enumerate}

\subsection{Bootstrap Confidence Bands}\label{sec_boot}
Let $\hat{\bSigma}_v^{(re)}$ be a regularized version of the sample covariance matrix $\hat{\bSigma}_v$, i.e., using regularization such as thresholding \citep{bickel2008covariance}, CLIME \citep{cai2011constrained}, LASSO Cholesky as in \citet{Margaritella17122024} or graphical LASSO \citep{meinshausen2006high,friedman2008sparse}. We use here the graphical LASSO which puts sparsity constraints on $\bSigma_v^{-1}$. The same estimator will be also used in Section \ref{sec_spect} to estimate the idiosyncratic spectral density matrix.

\begin{enumerate}[label=\it Step \arabic*:, leftmargin=*]
\item  \ Generate pseudo innovations
$\{\boeta_t^*, t \in \Z\},$ where $\boeta_t^*=((\BS u_t^*)^\top,(\BS v_t^*)^\top)^\top$ by drawing $\BS u^* \overset{i.i.d.}{\sim} \mathcal{N}(0,\hat\bSigma_u)$ and $\BS v^* \overset{i.i.d.}{\sim} \mathcal{N}(0,\hat\bSigma_v^{(re)})$.
\item  \  Generate a pseudo factor series  $ \{\BS f_t^*\}$ by using the VAR$(p_f)$ model equation, that is
$\bof_t^*=\sum_{j=1}^{p_f}\hat{\BS D}^{(j)} {\BS f^*}_{t-j}+\bu_t^*$ and a burn-in phase. Generate a pseudo idiosyncratic series  $ \{\BS \xi_t^*\}$ by using the VAR$(p_\xi)$ model equation, that is
$\BS \xi_t^*=\sum_{j=1}^{p_\xi}\hat{\BS B}^{(j)} {\BS \xi^*}_{t-j}+\bv_t^*$ and a burn-in phase. Use the factor model equation to generate a pseudo time series $\bx_t^*=\hat{\BS \Lambda}_i^\top \BS f_t^*+\BS \xi_t^*, t=1,2,\dots,T$.
\item \ Using the pseudo time series $\{\BS x_t^*\}$, estimate a factor models, i.e., the loadings, factors, and idiosyncratic part as in \eqref{eq.svd}. This gives $\hat {\BS f}_t^*, \hat {\BS \Lambda}^*, \hat {\BS \xi}_t^*$. Additionally, estimate a VAR model on the factors and a sparse VAR model on the idiosyncratics as described in step (1.) and (2.) of the previous algorithm. This leads to $\hat{\BS D^*}^{(j)},j=1,\dots,p_f,\hat{\BS B^*}^{(j)},j=1,\dots,p_\xi$.
\item \ Follow \cite{krampe2023structural} and compute the de-sparsified MA-matrices $\hat{\BS \psi}_\xi^{*,(de),(j)},j=1,\dots,H$ based on $\hat {\BS \xi}_t^*$ and the estimated sparse VAR $\hat{\BS B^*}^{(j)},j=1,\dots,p_\xi$. Additionally, estimate the variance of $\hat {\BS v_t}^*$ leading to $\hat{\BS \Sigma}_v^{*}$.
\item \ Using $\hat{\BS \psi}_f^{*,(j)},\hat{\BS \psi}_\xi^{*,(de),(j)}\; \text{for}\; j=1,\dots,H, \; \text{and}\; \hat{\BS \Sigma}_u^{*}, \hat{\BS \Sigma}_v^{*}$, compute the FEVD as in \eqref{theta} leading to $\hat\theta_{ij}^g(H)^*,i=1,\dots,N,j=1,\dots,N+r$. 
\item \ Approximate the distribution of $  \sqrt T (\hat\theta_{ij}^g(H)- \theta_{ij}^g(H))$ by the distribution  of the bootstrap analogue $ \sqrt T (\hat\theta_{ij}^g(H)^*- \hat\theta_{ij}^g(H))$ for $i=1,\dots,N,j=1,\dots,N+r$.
\end{enumerate}

\subsection{Details on Spectral Density Estimation - Frequency Domain}\label{sec_spect}
Let $K$ be a kernel estimator of the factors' sample periodogram, fulfilling the following two regularity assumptions\footnote{Let us note how absolute summability of the factors autocovariances is directly implied by Assumption \ref{ass.moments}} (same as Assumption~1, 2 in \citealp{wu2018asymptotic}):
\begin{assumption}\label{kernel_1}
 $K$ is an even and bounded function with bounded support in $(-1,1)$, continuous in $(-1,1)$, $K(0)=1, \int_{-1}^1 K^2(u) du<1$, and $\sum_{l \in \Z} \sup_{|s-l|<1} |K(l\omega)-K(s\omega)|=O(1)$ as $\omega \to 0$.   
\end{assumption}
\begin{assumption}\label{kernel_2}
There exist constants $0 < b_1 < b_2 < 1$ and $M_9,M_{10} > 0$ such that the lag-window size $B_T$ fulfills: $M_9 T^{b_1}\leq B_T \leq M_{10} T^{b_2}$, for all large $T$.
\end{assumption}
These requirements are quite general, as they hold for most of the commonly used kernels (e.g., the Barlett kernel).
Then, a spectral density estimator for the factors is given by 
\begin{align}\label{eq_est_spectdens_f}
    \hat{ \BS f}_f(\omega)=\frac{1}{2\pi}\sum_{h=-T+1}^{T-1} K\left(\frac{h}{B_T}\right) \exp(-\mathrm{i}h \omega) \hat {\BS \Gamma}_f(h),
\end{align}\normalsize
where  $\hat { \BS\Gamma}_f(h)=T^{-1}\sum_{t} \hat{\BS f}_{t+h} \hat{ \BS f}_t^{\top}$ is the sample autocovariance function. 
Consistency of $\hat {\BS f}_f(\omega)$ follows, as formalized in the following Lemma~\ref{lemma.spectral.density.factor} below. 

\begin{lemma}\label{lemma.spectral.density.factor}
Under Assumptions~\ref{ass.moments}-\ref{kernel_2}, consider the rotation matrix $\BS H_{NT}^\top=(\boldsymbol{\Lambda}^\top \boldsymbol{\Lambda}/N) (\boldsymbol{F}^\top \hat{\boldsymbol{F}}/T) \BS D_{NT,r}^{-2}$ and let \begin{align*}
g(N,T,\zeta)&=(NT)^{2/\zeta}\left(\frac{1}{\sqrt{N}T}+\frac{1}{T^{3/2}}+(NT)^{2/\zeta}\frac{1}{T^2}\right),
\end{align*} then,
\begin{align}\label{eq_spec_rate_factors}
    \| \boldsymbol{\hat{f}}_f(\omega)-\BS H_{NT}\BS f_f(\omega)\BS H_{NT}^\top\|_{\max}&=O_P\Bigg(\sqrt{B_T/T}+
    \frac{{\log(N)}}{{T}}+\frac{k_\xi}{N}+\frac{\sqrt{\log(N)}}{\sqrt{NT}}+\gt\Bigg)
\end{align}
\end{lemma}\normalsize
A word on the obtained estimation rate in \eqref{eq_spec_rate_factors}. 
In order to get consistency, it is unsurprisingly needed for both $N$ and $T$ to grow. Additionally, if $N = T^a$ and $0 \leq a \leq \zeta - 4$, we have $g(N, T, \zeta) \leq \frac{1}{\sqrt{NT}} + \frac{1}{T}$
which means $g(N, T, \zeta)$ could be dropped in $O_P$-notation. Likewise,  
$\frac{{\log(N)}}{{T}}$ and $\frac{\sqrt{\log(N)}}{\sqrt{NT}}$ can also be dropped if $N$ grows only polynomial with respect to $T$, which is a mild requirement. Hence, it remains $\frac{k_\xi}{N}$. First, note how the term $k_{\xi}$ from Assumption \ref{sparsity} quantifies the linear dependence of the idiosyncratic component. We follow \citet{krampe2021factor} who notes that $\sqrt{N}$ is an upper bound for the growth rate of $k_{\xi}$, and therefore a rate smaller than $\sqrt{N}$ is most sensible and in line with the factor models literature. Letting $N = T^a$ as above, we can then consider $k_{\xi}=O_P(T^{a/2-\varepsilon})$ for some small $\epsilon>0$. Substituting and simplifying one obtains $\frac{k_\xi \sqrt{T}}{N \sqrt{B_T}}=O_P(
T^{1/2-a/2-b_1-\eps})$. Thus, if $1/2\leq/2+b_1+\eps$ the estimation of the factors does not lead to a slower rate for the spectral density estimator.

Proof of Lemma \ref{lemma.spectral.density.factor}, given in the Appendix, hinges on the fact that the difference between the estimated spectral density $\boldsymbol{\hat{f}}_f(\omega)$, and the rotated version of the true one $\BS H_{NT}\BS f_f(\omega)\BS H_{NT}^\top$, can be bounded above by the difference between the former and an infeasible version of the former, plus the difference between the infeasible version and the true one. The infeasible version here contains in \eqref{eq_est_spectdens_f} $\tilde{\BS \Gamma}_{f} (h)=T^{-1}\sum_{t} \BS f_{t+h} \BS f_t^{\top}$, in place of $\hat {\BS \Gamma}_f(h)$, and results from \citet{wu2018asymptotic} and \citet{krampe2021factor} can then be straightforwadly applied to yield the consistency. Though it depends on the choice of the kernel, one would want to have an as small as possible bandwidth $B_T$, so as to approximate a parametric rate for $\norm{\boldsymbol{\tilde f}_f(\omega)-\BS f_f(\omega)}_{\max}$, while having an as smooth as possible spectra, i.e., large $q$ where $\lim_{x\to 0} \frac{1-K(x)}{|x|^q}<\infty$ \citep[see also][Remark (ii)]{wu2018asymptotic}.

Now onto the idiosyncratics. As mentioned, we use here the VAR structure of the idiosyncratic component to estimate its spectral density matrix. In Section \ref{sec_model}, the VAR parameters of the idiosyncratic component can be estimated row-wise consistently (see Assumption \ref{sparsity}), i.e., consistency of $\BS B$ for the matrix norm $\|\cdot \|_\infty$. However, the estimation of the spectral density requires additional column-wise consistency, that is consistency of $\BS B^{(j)},j=1,\dots,p_{\xi},$ with respect to $\|\cdot\|_1$. Such a column-wise consistency requires additional sparsity assumptions, see also \cite{krampe2020statistical} for a discussion. Furthermore, a parametric estimation of the spectral density matrix of a VAR process requires an estimate of the covariance or precision matrix of the residual process $\{\BS v_t\}$. 
Assumption~\ref{sparsity.b} below formalizes the additional sparsity assumption and the requirements on the residuals covariance matrix.
\begin{assumption}\label{sparsity.b}(\textit{Sparsity and stability})\\
$(i)$ 
The idiosyncratic VAR process is row- and column-wise approximately sparse with approximate sparsity parameter $q \in [0,1)$, i.e., 
$$\sum_{l=1}^{p_{\xi}} \max_i \sum_{j=1}^N |\BS B_{i,j}^{(l)}|^q\leq k, \qquad \sum_{l=1}^{p_{\xi}} \max_j \sum_{i=1}^N |\BS B_{i,j}^{(l)}|^q\leq  k.$$\normalsize
$(ii)$ As in Assumption~\ref{sparsity} (ii) and $\sup_\omega \|\BS f _\xi(\omega)\|_\infty \leq k_\xi M_9$.\\
$(iii)$ The covariance matrix $\BS \Sigma_v=\var(\BS v_t)^{-1}$ of the VAR innovations $\{\BS v_t\}$ is positive definite and approximately sparse and $\|\BS \Sigma_v\|_2\leq M_{10}$, where $M_{10}$ is a positive constant. Let $q_v \in[0,1)$ denote the approximate sparsity parameter and $k_v$ the sparsity. Then, 
$$
\max_i \sum_{j=1}^N |(\BS \Sigma_v)_{i,j}|^{q_v}=\max_j \sum_{i=1}^N |(\BS \Sigma_v)_{i,j}|^{q_v}\leq k_v. 
$$ \normalsize 
\end{assumption}

\begin{lemma} \label{lem.rate.spectral.var}
Under Assumption~\ref{ass.moments}, \ref{ass.fac}, \ref{sparsity.b}, for $l \in [1,\infty]$, and employing the Graphical Lasso \citep{meinshausen2006high} to get $\hat {\BS \Sigma}_v^{-1(re)}$, we then have the following:
\begin{align*}
    \| &\hat {\BS \Sigma}_v^{-1(re)}-{\BS \Sigma}_v^{-1}\|_l=O_P\Bigg(k_v \|{\BS \Sigma}_v^{-1}\|_1 \Bigg[\sqrt{(\log(N)/T})+N^{2/\zeta}T^{2/\zeta-1}+k\Bigg[\frac{k_\xi}{N}+\frac{\log(N)}{T}+\frac{\sqrt{\log(N)}}{\sqrt{NT}}\\
    &+(NT)^{2/\zeta-1}k_\xi+\frac{(NT)^{4/\zeta}}{T^2}\Bigg]+\left(\sqrt{(\log(Np)/T})+(Np)^{2/\zeta}T^{2/\zeta-1}\right)\Bigg(k\Bigg[\sqrt{\log(Np)/T}+(NpT)^{2/\zeta}/T\\
    &+k\Bigg(\frac{k_\xi}{N}+\frac{\sqrt{\log(Np)}}{\sqrt{NT}}+(NpT)^{2/\zeta}\Bigg(\frac{k_\xi}{NT}+\frac{1}{\sqrt{N}T}+\frac{1}{T^{3/2}}+(NpT)^{2/\zeta}\frac{1}{T^2}\Bigg)\Bigg)\Bigg]^{1-q}\Bigg)\Bigg]^{1-q_v}\Bigg),
\end{align*}

\begin{align*}
    &\|\BS f_\xi(\omega)^{-1}-\hat{\BS f}_\xi(\omega)^{-1}\|_l=O_P( k^2 (\|{\BS \Sigma}_v^{-1}-\hat {\BS \Sigma}_v^{-1,(re)}\|_l+\max_s  \|\hat \bbeta^{(s)}- \bbeta^{(s)}\|_2^{1-q} \|{\BS \Sigma}_v^{-1}\|_l ),\\
    &\|\BS f_\xi(\omega)^{-1}-\hat{\BS f}_\xi(\omega)^{-1}\|_2=O_P(\|{\BS \Sigma}_v^{-1}-\hat {\BS \Sigma}_v^{-1,(re)}\|_l+k\max_s  \|\hat \bbeta^{(s)}- \bbeta^{(s)}\|_2^{1-q} ).
\end{align*}
\normalsize

Now, to lighten the notation, let us further define the following quantities: $$C^*=\Bigg[\sqrt{(\log(N)/T})+k\Bigg[\frac{k_\xi}{N}+\frac{\log(N)}{T}+\frac{\sqrt{\log(N)}}{\sqrt{NT}}\Bigg]\Bigg],$$ $$D^*=\Big[\sqrt{\log(Np)/T}+k k_\xi/N+k \sqrt{\log(Np)/(NT)}\Big],$$ $$E^*=C^*-k \frac{\log(N)}{T}=\Bigg[\sqrt{\log(N)/T}+k k_\xi/N+k \frac{\sqrt{\log(N)}}{\sqrt{NT}}\Bigg].$$

Then, if $N=T^a, p=T^b$ for some $a,b>0$, $\zeta\geq 4(1+a+b)$ and $k$ such that $\|\hat \A-\A\|_\infty=o_P(1)$, these error bounds simplify to 
\begin{align*}
    \| \hat {\BS \Sigma}_v^{-1,(re)}-{\BS \Sigma}_v^{-1}\|_l&=O_P\Bigg(k_v \|{\BS \Sigma}_v^{-1}\|_1 \left[C^*\right]^{1-q_v}\Bigg),\\
    \|\BS f_\xi(\omega)^{-1}-\hat{\BS f}_\xi(\omega)^{-1}\|_l&=O_P\Bigg(k^2 \|{\BS \Sigma}_v^{-1}\|_1 \Big(k_v  \left[C^*\right]^{1-q_v} +\sqrt{k} \left[D^*\right]^{1-q/2} \Big)\Bigg),\\
    \|\BS f_\xi(\omega)^{-1}-\hat{\BS f}_\xi(\omega)^{-1}\|_2&=O_P\Bigg(k_v \|{\BS \Sigma}_v^{-1}\|_1 \left[C^*\right]^{1-q_v}+k^{3/2} \left[D^*\right]^{1-q/2} \Bigg).
\end{align*}
\end{lemma}
Now, we can give the error bound for the estimator of the spectral density of the whole process $\bof_{x}(\omega)$.  
We only present here explicitly the rate for a simplified case. In the general case, an explicit rate can be obtained by inserting the results of Lemma~\ref{lem.rate.spectral.var} and Theorem 1 of \citet{krampe2021factor}. Since it leads to a lengthy and not insightful expression, we omit it here. The rate is dominated by the estimation error of the sparse VAR and it is similar to the one in Theorem 1 of \citet{krampe2021factor}. However, the rate is more affected by the sparsity parameter in the sense that its maximum growth rate is less for the spectral density than it is for prediction. Maximum growth rate refers here to the maximal rate of sparsity for which consistency can be achieved. 

\begin{theorem}\label{thm.spectral.density}
Under Assumptions~\ref{ass.moments}-\ref{sparsity.b}, for $l \in [1,\infty]$ we  have the following 
\begin{align*}
    \|\BS f_x(\omega)^{-1}-\hat{\BS f}_x(\omega)^{-1}\|_l&=O_P(k_\xi\|\hat{\BS f}_\xi^{-1}(\omega)-{\BS f}_\xi^{-1}(\omega)\|_\infty+k_\xi^2 \|\hat {\BS  \Lambda}-\BS \Lambda \bH_{NT}^{-1}\|_{\max}),\\ 
    \|\BS f_x(\omega)^{-1}-\hat {\BS f}_x(\omega)^{-1}\|_2&=O_P(\|\hat{\BS f}_\xi^{-1}(\omega)-{\BS f}_\xi^{-1}(\omega)\|_2+\|\hat {\BS \Lambda}-\BS \Lambda \bH_{NT}^{-1}\|_{\max}).
\end{align*}
\normalsize
If $N=T^a, p=T^b$ for some $a,b>0$, $\zeta\geq 4(1+a+b)$ and $k=o(\sqrt{T/\log(Np)})$, these error bounds simplify to 
\begin{align*}
    \|\BS f_x(\omega)^{-1}-\hat {\BS f}_x(\omega)^{-1}\|_l&=O_P\Bigg( k^2 \|{\BS \Sigma}_v^{-1}\|_1 \Big(k_v  \left[C^*\right]^{1-q_v} +\sqrt{k} \left[E^*\right]^{1-q/2} \Big)\Bigg),\\
    \|\BS f_x(\omega)^{-1}-\hat {\BS f}_x(\omega)^{-1}\|_2&=O_P\Bigg(k_v \|{\BS \Sigma}_v^{-1}\|_1 \left[C^*\right]^{1-q_v}+k^{3/2} \left[E^*\right]^{1-q/2} \Bigg).
\end{align*}
\end{theorem}\normalsize

\section{Conclusion}\label{sec_conclusion}
We decompose the high-dimensional global bank network connectedness index into connectedness due to market shocks, idiosyncratic shocks, and shocks at high, medium, and low frequencies. Instead of regularizing the high-dimensional vector of banks with sparsity-inducing estimators, we use recent literature linking factor models with sparse ones. We estimate a static, approximate factor model with sparse VAR idiosyncratic components, enabling decomposition of connectedness into these parts and providing bootstrap confidence bands. We also analyze the spectral counterpart to disentangle frequency responses to shocks. Our findings show that idiosyncratic variation largely drives the highly interconnected network of bank stock price volatilities, especially during non-turbulent periods and in the long run. However, during major crises like the 2008 financial crisis and Covid-19, bank stock volatilities become more interconnected, with connections driven by short-run market dynamics. \newpage

\bibliography{references}
\newpage
\begin{appendix}

\section{Proofs}

\begin{proof}[Proof of Lemma~\ref{lemma.spectral.density.factor}]
We have \begin{align*}\|\boldsymbol{\hat f}_f(\omega)-\BS H_{NT}\BS f_f(\omega)\BS H_{NT}^\top\|_{\max}&\leq\underbrace{\|\boldsymbol{\hat f}_f(\omega)-\BS H_{NT}\boldsymbol{\tilde f}_f(\omega)\BS H_{NT}^\top\|_{\max}}_{(i)}\\&+\underbrace{\|\boldsymbol{\tilde f}_f(\omega)-\BS f_f(\omega)\|_{\max}\|\BS H_{NT}\|_1 \|\BS H_{NT}\|_\infty}_{(ii)},\end{align*} 
which follows by adding and subtracting $\BS H_{NT} \boldsymbol{\tilde f}_f(\omega)\BS H_{NT}^{\top}$, triangle inequality, submultiplicativity of the matrix norm, the fact that $\norm{\bA}_{\max}\leq \norm{\bA}_1$ for any matrix $\bA$, and finally Hölder inequality.
For the term $(ii)$, we have by Lemma A1.1 in \citet{krampe2021factor} that $\|\BS H_{NT}\|_l=O(1)$ for $l\in\{1,\infty\}$. Furthermore, by Theorem 3 in \cite{wu2018asymptotic} $\|\boldsymbol{\tilde f}_f(\omega)-\BS f_f(\omega)\|_{\max}=O_P(\sqrt{B_T/T})$. Now for term $(i)$. Note first that the dimension $r$ of the process $\{\BS f_t\}$ is fixed. Then, we have 
\begin{align*}&(i)=\left\|\frac{1}{2\pi} \sum_{h=-T+1}^{T-1} K\left(\frac{h}{B_T}\right) \exp(-\mathrm{i}h \omega) \underbrace{(\boldsymbol{\hat \Gamma}_f(h)-\BS H_{NT}  \boldsymbol{\tilde{\Gamma}}_f(h) \BS H_{NT}^\top)}_{(iii)}\right\|_{\max}.\end{align*} Then, we can rewrite $(iii)$ by adding an subtracting $\BS  H_{NT}\boldsymbol{f}_{t}$ and $\BS H_{NT}\boldsymbol{f}_{t+h}$ as follows

\begin{align*}
 (iii)=&T^{-1}\sum_t \boldsymbol{\hat f}_{t+h}\boldsymbol{\hat f}_{t}-\BS H_{NT}(T^{-1}\sum_t \boldsymbol{f}_{t+h}\boldsymbol{f}_{t}^\top)\BS H_{NT}^{\top}\\
 =&T^{-1}\sum_t \BS H_{NT}\boldsymbol{f}_{t+h})(\boldsymbol{\hat f}_{t}-\BS H_{NT}\boldsymbol{f}_{t})^{\top}+\\
 &T^{-1}\sum_t (\boldsymbol{\hat f}_{t+h}-\BS H_{NT}\boldsymbol{f}_{t+h})(\BS H_{NT}\boldsymbol{f}_{t})^{\top}+\\
 &T^{-1}\sum_t (\boldsymbol{\hat f}_{t+h}-\BS H_{NT}\boldsymbol{f}_{t+h})(\boldsymbol{\hat f}_{t}-\BS H_{NT}\boldsymbol{f}_{t})^{\top}.
\end{align*}

Furthermore, note that we have by Lemma~1 H) in \cite{krampe2021factor} for $t\in \Z$
\begin{align*}
    (\hat{\boldsymbol{f_t}}-\BS H_{NT}\boldsymbol{f_t})^\top=&\frac{1}{NT}\left[\sum_{i=1}^N \sum_{s=1}^T \xi_{i,t} \BS\Lambda_i^\top \BS f_{s} {\BS f_s}^\top+ \sum_{i=1}^N \sum_{s=1}^T \xi_{i,t}\xi_{i,s} {\BS f_s^\top}\right] \BS H_{NT}^\top \BS D_{NT,r}^{-2}\\
    &+O_P\left(\frac{{\log(N)}}{{T}}+\frac{k_\xi}{N}+\frac{\sqrt{\log(N)}}{\sqrt{NT}}+\gt\right). 
\end{align*}

Hence, this can be inserted for each of the differences. We can follow the arguments of the proof of Lemma~1 I) in \cite{krampe2021factor} and note that $\max_j\frac{1}{2\pi} \sum_{h=-T+1}^{T-1} K\left(\frac{h}{B_T}\right) \exp(-\mathrm{i}h \omega) e_j\BS H_{NT}\boldsymbol{f_t}\xi_{i,t}=O_P(1)$. This results in 
$$
(i)=O_P\left(\frac{{\log(N)}}{{T}}+\frac{k_\xi}{N}+\frac{\sqrt{\log(N)}}{\sqrt{NT}}+\gt\right).
$$

\end{proof}
\begin{proof}[Proof of Lemma~\ref{lem.rate.spectral.var}]
First, let us consider the estimation error in the residuals. For this, we consider the (unfeasible) sample covariance $\tilde {\BS \Sigma}_v=T^{-1} \sum_t \BS v_t \BS v_t^\top$. Based on $\zeta$ moments (see Assumption \ref{ass.moments}) and a Nagaev's inequality for dependent processes (see Section 2.1 in \citealp{wu2016performance}),
$\|\tilde {\BS \Sigma}_v-\BS \Sigma_v\|_{\max}=O_P(\sqrt{(\log(N)/T})+N^{2/\zeta}T^{2/\zeta-1})$. Note that we have only the estimated residuals, given by $\hat{\BS v}_t=\hat{\BS \xi}_t-\sum_{j=1}^{p_{\xi}} \hat{\BS B}^{(j)} \hat{\BS \xi}_{t-j}$. This gives the sample covariance 
$\hat {\BS \Sigma}_v=T^{-1} \sum_t \hat{\BS v}_t \hat {\BS v}_t^{\top}$. We have 
$$\tilde {\BS \Sigma}_v- \hat {\BS \Sigma}_v=T^{-1} \sum_t ({\BS v}_t-\hat{\BS v}_t) {\BS v}_t^{\top}+ \BS v_t ({\BS v}_t-\hat{\BS v}_t)^{\top} +(\hat{\BS v}_t-{\BS v}_t)({\BS v}_t-\hat{\BS v}_t)^{\top}.$$



Furthermore, as ${\BS v}_t={\BS \xi}_t-\sum_{j=1}^{p_{\xi}} {\BS B}^{(j)} {\BS \xi}_{t-j}$ and $\hat{\BS v}_t=\hat{\BS \xi}_t-\sum_{j=1}^{p_{\xi}} \hat{\BS B}^{(j)} \hat{\BS \xi}_{t-j}$, for $\BS w_t:=\BS {\hat \xi_t}-\BS \xi_t$ we can rewrite, $${\BS v}_t-\hat{\BS v}_t={\BS w}_t+\sum_{j=1}^{p_{\xi}} \BS B^{(j)} \BS w_{t-j}+\sum_{j=1}^{p_{\xi}}(\hat{\BS B}^{(j)}-\BS B^{(j)}) \BS \xi_{t-j}+\sum_{j=1}^{p_{\xi}} (\hat{\BS B}^{(j)}-\BS B^{(j)}) \BS w_{t-j}.$$ Hence, using the stacked version of the VAR matrix, Hölder's inequality, and following the arguments of Theorem 1 in \citet{krampe2021factor}, we have \begin{align*}
    \|\tilde {\BS \Sigma}_v- \hat {\BS \Sigma}_v\|_{\max}&=O_P\Big(\|\A\|_\infty \|T^{-1} \sum_{t} \BS w_t \BS \xi_t\|_{\max}+ \\&+\max_j \|\hat \bbeta^{(j)}- \bbeta^{(j)}\|_1 \Big( \|T^{-1}\sum_t \BS \xi_{t-1}^v \BS v_t\|_{\max}+\|T^{-1} \sum_{t=1} \BS w_t \BS \xi_t\|_{\max}\Big)\Big).
\end{align*} 
Since $\mathbb{E}(\BS \xi_{t-1}^v \BS v_t)=0$, we have by the arguments of Lemma A1.1 in \citet{krampe2021factor} $\|T^{-1}\sum_t \BS \xi_{t-1}^v \BS v_t\|_{\max}=O_P(\sqrt{(\log(Np)/T})+(Np)^{2/\zeta}T^{2/\zeta-1})$. Together with Theorem 1 in \citet{krampe2021factor} and $\|\tilde {\BS \Sigma}_v-\BS \Sigma_v\|_{\max}$  this lead to the following rate:
\begin{align*}
&\|{\BS \Sigma}_v- \hat {\BS \Sigma}_v\|_{\max}=\|{\BS \Sigma}_v-\tilde {\BS \Sigma}_v+\tilde {\BS \Sigma}_v- \hat {\BS \Sigma}_v\|_{\max} \\&=O_P\Bigg(\sqrt{(\log(N)/T})+N^{2/\zeta}T^{2/\zeta-1}+k\Bigg[\frac{k_\xi}{N}+\frac{\log(N)}{T}+\frac{\sqrt{\log(N)}}{\sqrt{NT}}+(NT)^{2/\zeta-1}k_\xi+\frac{(NT)^{4/\zeta}}{T^2}\Bigg]+ \\
&\left(\sqrt{(\log(Np)/T})+(Np)^{2/\zeta}T^{2/\zeta-1}\right)\Bigg(k\Bigg[\sqrt{\log(Np)/T}+(NpT)^{2/\zeta}/T+k\Bigg(\frac{k_\xi}{N}+\frac{\sqrt{\log(Np)}}{\sqrt{NT}}\\
&+(NpT)^{2/\zeta}\Bigg(\frac{k_\xi}{NT}+\frac{1}{\sqrt{N}T}+\frac{1}{T^{3/2}}+(NpT)^{2/\zeta}\frac{1}{T^2}\Bigg)\Bigg)\Bigg]^{1-q}\Bigg). \end{align*}
Using Graphical Lasso (or CLIME) to get $\{\hat{\BS v}_t\}$ leads to a plug-in precision matrix estimator as $\hat {\BS \Sigma}_v^{-1(re)}$; now following the arguments of \citet{meinshausen2006high} (or \citealp{cai2011constrained}) gives us that the Graphical Lasso (or CLIME) estimator fulfills $\|{\BS \Sigma}_v^{-1}-\hat {\BS \Sigma}_v^{-1(re)}\|_l=O_P(k_v (\|{\BS \Sigma}_v^{-1}\|_1 \|{\BS \Sigma}_v- \hat {\BS \Sigma}_v\|_{\max}^{1-q_v}))$ for $l\in [1,\infty]$. We have by Theorem~2 in \citet{krampe2021factor} that $\|\A-\hat{\A}\|_{\max}=O_P( \max_s\|\hat \bbeta^{(s)}- \bbeta^{(s)}\|_2)$. Consequently, we obtain by Theorem 1 in \cite{krampe2020statistical} that under Assumption~\ref{sparsity.b} $\sum_{j=1}^{p_{\xi}} \|\hat{\BS B}^{(thr,j)}-\BS B^{(j)}\|_l= O(k \max_s\|\hat \bbeta^{(s)}- \bbeta^{(s)}\|_2^{1-q})$. Then, we have by Theorem~6 in \cite{krampe2020statistical}
\begin{align*}\|\BS f_\xi(\omega)^{-1}-\hat{\BS f}_\xi(\omega)^{-1}\|_l=O_P\Bigg(\sum_{j=1}^{p_{\xi}}\|\BS B^{(j)}\|_l^2 \|{\BS \Sigma}_v^{-1}-\hat {\BS \Sigma}_v^{-1(re)}\|_l+\sum_{j=1}^{p_{\xi}} \|\hat{\BS B}^{(thr,j)}-\BS B^{(j)}\|_l \|\BS B^{(j)}\|_l \|\BS \Sigma_v\|_l\Bigg).\end{align*} 
\end{proof}
\begin{proof}[Proof of Theorem~\ref{thm.spectral.density}]
We have:
\begin{align}
    &\|\boldsymbol{f}_x(\omega)^{-1}-\boldsymbol{\hat f}_x(\omega)^{-1}\|_l\leq \|\BS f_\xi^{-1}(\omega)-\hat{\BS f}_\xi^{-1}(\omega)\|_l\label{eq.spec.DFM}\\
&+\Big\|\hat{\BS f}_\xi^{-1}(\omega) \boldsymbol{\hat \Lambda} \Big(\boldsymbol{\hat f}_f^{-1}(\omega)/N+\boldsymbol{\hat \Lambda}^\top/\sqrt{N} \hat{\BS f}_\xi^{-1}(\omega)\boldsymbol{ \hat \Lambda}/\sqrt{N}\Big)^{-1} \boldsymbol{\hat \Lambda}^\top/N \hat{\BS f}_\xi^{-1}(\omega) \nonumber\\&-\BS f_\xi^{-1}(\omega) \BS \Lambda \BS H_{NT}^{-1} \Big((\BS H_{NT}^{-1})^\top \boldsymbol{f}_f^{-1}(\omega)\BS H_{NT}^{-1}/N+(\BS H_{NT}^{-1})^\top\BS \Lambda^\top /\sqrt{N}\BS f_\xi^{-1}(\omega) \BS \Lambda \BS H_{NT}^{-1}/\sqrt{N}\Big)^{-1}\nonumber\\&\times (\BS H_{NT}^{-1})^\top \BS \Lambda^\top/N \BS f_\xi^{-1}(\omega)
\Big\|_l \nonumber, 
\end{align}\normalsize
Let $G=((\BS H_{NT}^{-1})^\top  \BS f_f^{-1}(\omega)\BS H_{NT}^{-1}/N+(\BS H_{NT}^{-1})^\top\BS \Lambda^\top \BS f_\xi^{-1}(\omega) \BS \Lambda \BS H_{NT}^{-1}/N)^{-1}$\normalsize and $\hat G=(\boldsymbol{\hat f}_f^{-1}(\omega)/N+\boldsymbol{\hat  \Lambda}^\top \hat{\BS f}_\xi^{-1}(\omega) \boldsymbol{\hat \Lambda})^{-1}/N$. \normalsize Lemma~\ref{lem.rate.spectral.var} gives a rate for $\|\BS f_\xi^{-1}(\omega)-\hat{\BS f}_\xi^{-1}(\omega)\|_l$. Furthermore, the second term on the right hand side of \eqref{eq.spec.DFM} is smaller or equal to:
\begin{align*}
&\|\hat{\BS f}_\xi^{-1}(\omega) \boldsymbol{\hat \Lambda}-{\BS f}_\xi^{-1}(\omega) \BS \Lambda (\BS H_{NT}^{-1})\|_l \| G\|_l \|(\BS H_{NT}^{-1})^\top \BS \Lambda^\top/N \BS f_\xi^{-1}(\omega)\|_l\\& + \| \BS f_\xi^{-1}(\omega) \BS \Lambda \BS H_{NT}^{-1}\|_l  \|G-\hat G\|_l \|(\BS H_{NT}^{-1})^\top \BS \Lambda^\top/N \BS f_\xi^{-1}(\omega)\|_l \\
&+ \| \BS f_\xi^{-1}(\omega) \BS \Lambda \BS H_{NT}^{-1}\|_l  \|G\|_l \| \boldsymbol{\hat \Lambda}^\top/N \hat{\BS f}_\xi^{-1}(\omega)-(\BS H_{NT}^{-1})^\top \BS \Lambda^\top/N \BS f_\xi^{-1}(\omega)\|_l \\
&+
\|\hat{\BS f}_\xi^{-1}(\omega) \boldsymbol{\hat \Lambda}-{\BS f}_\xi^{-1}(\omega) \BS \Lambda (\BS H_{NT}^{-1})\|_l \|G-\hat G\|_l \|(\BS H_{NT}^{-1})^\top \BS \Lambda^\top/N \BS f_\xi^{-1}(\omega)\|_l\\
&+\| \BS f_\xi^{-1}(\omega) \BS \Lambda \BS H_{NT}^{-1}\|_l \|G-\hat G\|_l \| \boldsymbol{\hat \Lambda}^\top/N \hat{\BS f}_\xi^{-1}(\omega)-(\BS H_{NT}^{-1})^\top \BS \Lambda^\top/N \BS f_\xi^{-1}(\omega)\|_l\\
&+\|\hat{\BS f}_\xi^{-1}(\omega) \boldsymbol{\hat \Lambda}-{\BS f}_\xi^{-1}(\omega) \BS \Lambda (\BS H_{NT}^{-1})\|_l \| G\|_l \| \boldsymbol{\hat \Lambda}^\top/N \hat{\BS f}_\xi^{-1}(\omega)-(\BS H_{NT}^{-1})^\top \BS \Lambda^\top/N \BS f_\xi^{-1}(\omega)\|_l\\
&+\|\hat{\BS f}_\xi^{-1}(\omega) \boldsymbol{\hat \Lambda}-{\BS f}_\xi^{-1}(\omega) \BS \Lambda (\BS H_{NT}^{-1})\|_l \|G-\hat G\|_l \| \boldsymbol{\hat \Lambda}^\top/N \hat{\BS f}_\xi^{-1}(\omega)-(\BS H_{NT}^{-1})^\top \BS \Lambda^\top/N \BS f_\xi^{-1}(\omega)\|_l,
\end{align*}\normalsize
$G$ is of fixed dimension $r\times r$ and we first show that $\|G\|_l=O(1),\; l \in [1,\infty]$. For this, we have $$\|G\|_2\leq \Big(\sigma_{\min}((\BS H_{NT}^{-1})^\top \BS f_f^{-1}(\omega)\BS H_{NT}^{-1}/N)+\sigma_{\min}((\BS H_{NT}^{-1})^\top\BS \Lambda^\top \BS f_\xi^{-1}(\omega) \BS \Lambda \BS H_{NT}^{-1}/N)\Big)^{-1}.$$\normalsize Note that Lemma~A1.1, A) in \citet{krampe2021factor} implies that $1/\sigma_{\min}(\BS H_{NT})=O(1)$ and $1/\sigma_{\min}(\BS H_{NT}^{-1})=O(1)$ and we have for symmetric matrices $A,B$, $1/\sigma_{\min}(AB)\leq 1/(\sigma_{\min} (A) \sigma_{\min}(B))$. Hence, $\sigma_{\min}((\BS H_{NT}^{-1})^\top \BS f_f^{-1}(\omega)\BS H_{NT}^{-1}/N)=O(1/N)$. Furthermore, let $\tilde {\BS \Lambda}=(\BS \Lambda^\top \BS \Lambda/N)^{-1/2} \BS \Lambda$. Note that $\BS \Lambda^\top \BS \Lambda/N=\BS \Sigma_\Lambda+o(1)$ and $\BS \Sigma_\Lambda$ is positive definite by Assumption~\ref{ass.fac} and also  $\sigma_{\min}(\BS \Lambda^\top \BS \Lambda/N)>1/M>0$. Then, $\tilde {\BS \Lambda}^\top \tilde {\BS \Lambda}/N=I_r$ and we have by Poincare's separation theorem $$\sigma_{\min}((\BS H_{NT}^{-1})^\top\BS \Lambda^\top \BS f_\xi^{-1}(\omega) \BS \Lambda \BS H_{NT}^{-1}/N)\geq \sigma_{\min}(\BS H_{NT}^{-1})^2 \sigma_{\min}((\BS \Lambda^\top \BS \Lambda/N)^{-1}) \sigma_{\min}(\BS f_\xi^{-1}(\omega)).$$ Thus, $\|G\|_2=O(1)$ and since it is of fixed dimension, we also have $\|G\|_l=O(1), l \in [1,\infty]$.
Since $\BS f_x$ is hermitian, we can focus on $l=\infty$. We have by Assumption~\ref{ass.fac} and \ref{sparsity.b} $$\| \BS f_\xi^{-1}(\omega) \BS \Lambda \BS H_{NT}^{-1}\|_\infty \leq \| \BS f_\xi^{-1}(\omega) \|_\infty \| \BS \Lambda \|_\infty \|\BS H_{NT}^{-1}\|_\infty \leq O(k_\xi).$$ Note that $\BS \Lambda \in {N\times r}$, which means $\| \BS \Lambda \|_\infty\leq r \|\BS \Lambda\|_{\max}=O(1)$.\\

Similarly, since $\|\BS \Lambda^\top/N\|_\infty \leq N/N \|\BS \Lambda\|_{\max}$ \normalsize, we have $\|(\BS H_{NT}^{-1})^\top \BS \Lambda^\top/N \BS f_\xi^{-1}(\omega)\|_\infty=O(k_\xi)$. By similar arguments, we have \begin{align*}&\|\hat{\BS f}_\xi^{-1}(\omega) \boldsymbol{\hat\Lambda}-{\BS f}_\xi^{-1}(\omega) \BS \Lambda (\BS H_{NT}^{-1})\|_\infty=\\&=O_P(\|\hat{\BS f}_\xi^{-1}(\omega)-{\BS f}_\xi^{-1}(\omega)\|_\infty+k_\xi \|\boldsymbol{\hat\Lambda}-\BS \Lambda \BS H_{NT}^{-1}\|_{\max})\\&=O_P(\|\hat{\BS f}_\xi^{-1}(\omega)-{\BS f}_\xi^{-1}(\omega)\|_\infty+k_\xi(\sqrt{\log(N)/T}+(NT)^{2/\zeta}/T+k_\xi/N)),\end{align*}\normalsize and \begin{align*}&\| \boldsymbol{\hat\Lambda}^\top/N \hat{\BS f}_\xi^{-1}(\omega)-(\BS H_{NT}^{-1})^\top \BS \Lambda^\top/N \BS f_\xi^{-1}(\omega)\|_\infty=\\&=O_P(\|\hat{\BS f}_\xi^{-1}(\omega)-{\BS f}_\xi^{-1}(\omega)\|_\infty+k_\xi(\sqrt{\log(N)/T}+(NT)^{2/\zeta}/T+k_\xi/N)).\end{align*}

We have further $\|G-\hat G\|_2 \leq \|G\|_2 \|\hat G\|_2 \|G^{-1}-\hat G^{-1}\|_2$ and  \begin{align*}
    \|G^{-1}-\hat G^{-1}\|_2&\leq \|(\BS H_{NT}^{-1})^\top \BS f_f^{-1}(\omega)\BS H_{NT}^{-1}/N-\hat{\BS f}_f^{-1}(\omega)/N\|_2\\&+\|(\BS H_{NT}^{-1}/\sqrt{N})^\top\BS \Lambda^\top \BS f_\xi^{-1}(\omega) \BS \Lambda \BS H_{NT}^{-1}/\sqrt{N}-\boldsymbol{\hat\Lambda}^\top/\sqrt{N} \hat{\BS f}_\xi^{-1}(\omega) \boldsymbol{\hat\Lambda})^{-1}/\sqrt{N}\|_2.
\end{align*}
 \normalsize Note \begin{align*}\|(\BS H_{NT}^{-1}/\sqrt{N})^\top\BS \Lambda^\top-\boldsymbol{\hat\Lambda}^\top/\sqrt{N}\|_2&\leq \|(\BS H_{NT}^{-1})^\top\BS \Lambda^\top-\boldsymbol{\hat\Lambda}^\top\|_{\max}\\&=O_P(\sqrt{\log(N)/T}+(NT)^{2/\zeta}/T+k_\xi/N),\end{align*} $\|(\BS H_{NT}^{-1}/\sqrt{N})^\top\BS \Lambda^\top\|_2=O(1)$ and $\|\BS f_\xi^{-1} (\omega)\|_2=O(1)$. \normalsize\\ 

Hence, by these results and Lemma~\ref{lemma.spectral.density.factor} we have \begin{align*}\|G^{-1}-\hat G^{-1}\|_2&=O_P\Big(\sqrt{\log(N)/T}+(NT)^{2/\zeta}/T+k_\xi/N+ \|\BS f_\xi(\omega)^{-1}-\hat{\BS f}_\xi(\omega)^{-1}\|_2\Big)\\&=O_P\Big(\sqrt{\log(N)/T}+(NT)^{2/\zeta}/T+k_\xi/N+\|{\BS \Sigma}_v^{-1}-\hat {\BS \Sigma}_v^{-1,(re)}\|_2\\&+k\max_s \|\hat \bbeta^{(s)}- \bbeta^{(s)}\|_2^{1-q}\Big)\end{align*} which is faster than $\|\hat{\BS f}_\xi^{-1}(\omega) \boldsymbol{\hat\Lambda}-{\BS f}_\xi^{-1}(\omega) \BS \Lambda (\BS H_{NT}^{-1})\|_\infty$. That means 
$$\|\BS f_x(\omega)^{-1}-\boldsymbol{\hat f}_x(\omega)^{-1}\|_\infty=O_P(k_\xi\|\hat{\BS f}_\xi^{-1}(\omega)-{\BS f}_\xi^{-1}(\omega)\|_\infty+k_\xi^2 \|\boldsymbol{\hat\Lambda}-\BS \Lambda \BS H_{NT}^{-1}\|_{\max}).$$
Since $\| \BS \Lambda \BS H_{NT}^{-1}/\sqrt{N}\|_2=O(1)$ and $\|\BS f_\xi(\omega)\|_2=O(1)$, we have further $$\|\BS f_x(\omega)^{-1}-\boldsymbol{\hat f}_x(\omega)^{-1}\|_2=O_P(\|\hat{\BS f}_\xi^{-1}(\omega)-{\BS f}_\xi^{-1}(\omega)\|_2+\|\boldsymbol{\hat\Lambda}-\BS \Lambda \BS H_{NT}^{-1}\|_{\max}).$$ The assertions follows after inserting the rates of Lemma~\ref{lem.rate.spectral.var}. 
\end{proof}

\clearpage
\section{Additional Figures}
\label{FiguresAppendix}

\begin{figure}[htbp]
    \centering
\includegraphics[page=2, width=0.95\textwidth, height=4.5cm]{Result_up_to_2013_2_spec.pdf}
\includegraphics[page=1, width=0.95\textwidth, height=6.5cm]{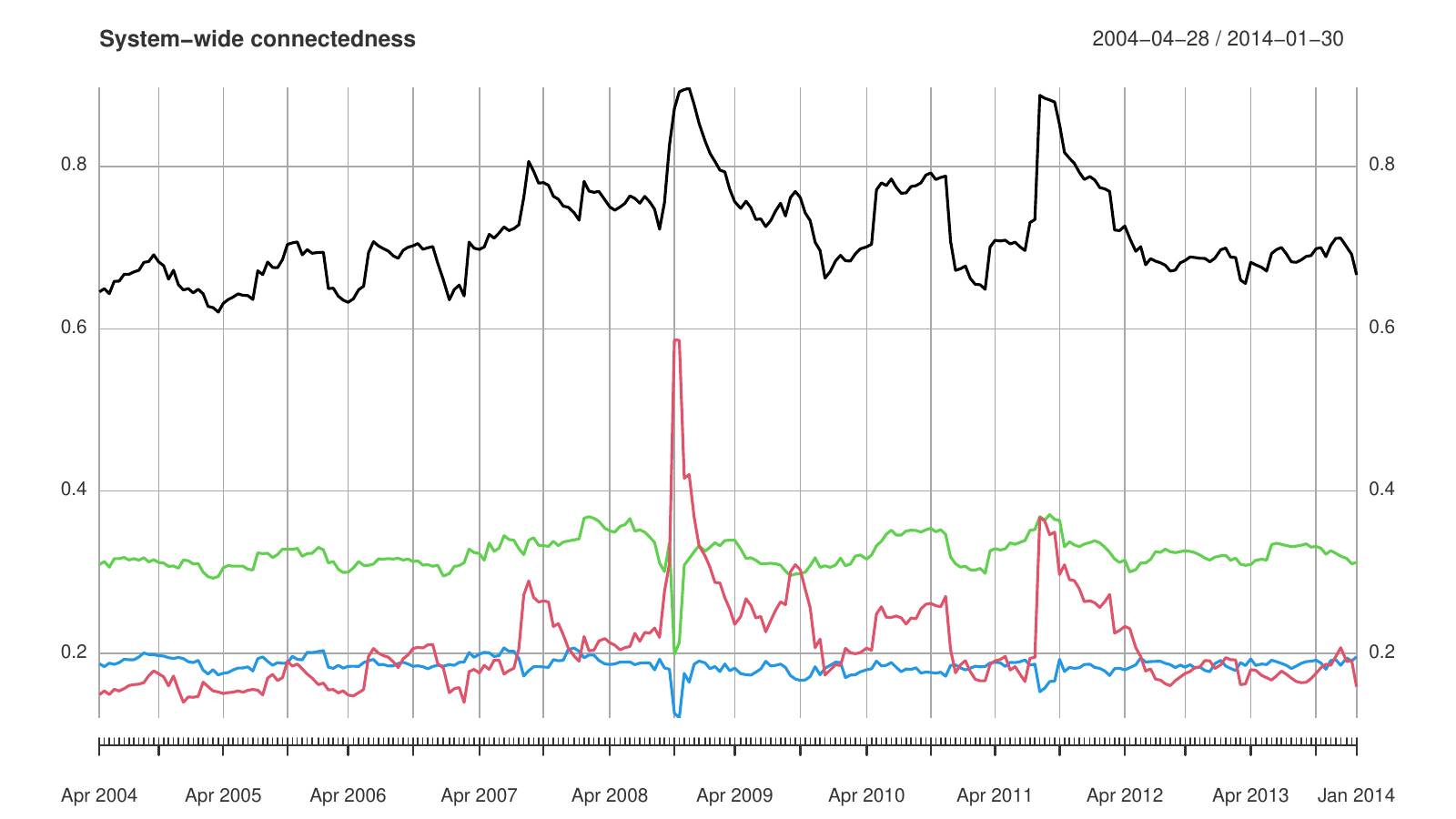} 
\includegraphics[page=3, width=0.95\textwidth, height=6.5cm]{Result_up_to_2013_2_spec.pdf}\caption{Top panel: System-Wide Connectedness ($C^H$); Center panel: System-Wide Connectedness due to common component ($C^H_{Mkt}$); Bottom panel: System-Wide Connectedness due to Idiosyncratics ($C^H_{Ids}$). Span: 2003-2013, 150 days rolling window.}\label{figure_app1}
\end{figure}
\clearpage
\section{Global Bank Detail, by Assets}
\label{DataAppendix}

Here we provide detail on our sample of all 96 publicly-traded banks in the world's top 150 (by assets). In Table \ref{Bank List} we show the banks ordered by assets, and we provide market capitalizations and assets (both in billions of U.S. dollars), our bank codes (which shows country), and Reuters tickers. The last column indicates with a tick if the bank is included in the new dataset from 2014 till 2023, or with a cross if not. 
\setcounter{table}{0} 
\renewcommand*{\thetable}{A\arabic{table}}
\singlespace
{\scriptsize
\begin{center}
\begin{longtable}[ht]{lcccccc} 
\caption{Global Bank Detail (Ordered by Assets)}  \\ \hline \hline
{\bf Bank }&    {\bf Country}   &       {\bf Mcap} &    {\bf Asset} &   {\bf Bank} & {\bf Reuters } & {\bf Dataset (ii) }  \\
{\bf Name }             &       &       &       &       {\bf Code} &    {\bf Ticker} \\      \hline
\endfirsthead
\multicolumn{4}{c}%
{\tablename\ \thetable\ -- \textit{Continued from previous page}} \\
\hline
{\bf Bank }&    {\bf Country}   &       {\bf Mcap} &    {\bf Asset} &   {\bf Bank} & {\bf Reuters } & {\bf Dataset (ii) }  \\
{\bf Name }                     &       &       &       &       {\bf Code} &       {\bf Ticker}            \\      \hline
\endhead
\hline \multicolumn{4}{r}{\textit{Continued on next page}} \\
\endfoot
\hline \hline
\endlastfoot
HSBC Holdings & UK & 2010 & 2.671 & hsba.gb & hsba.ln & \checkmark \\       
Mitsubishi UFJ Financial Group & Japan & 822 & 2.504 & mtbh.jp & X8306.to& \checkmark \\       
BNP Paribas & France & 1000 & 2.482 & bnp.fr & bnp.fr& \checkmark \\       
JPMorgan Chase \& Co & US & 2180 & 2.416 & jpm.us & jpm& \checkmark \\       
Deutsche Bank & Germany & 498 & 2.224 & dbk.de & dbk.xe& \checkmark \\ \hline
Barclays & UK & 682 & 2.174 & barc.gb & barc.ln& \checkmark \\ 
Credit Agricole & France & 367 & 2.119 & aca.fr & aca.fr& \checkmark \\ 
Bank of America & US & 1770 & 2.102 & bac.us & bac& \checkmark \\ 
Citigroup & US & 1500 & 1.880 & c.us & c& \checkmark \\ 
Mizuho Financial Group & Japan & 497 & 1.706 & mzh.jp & 8411.to& \checkmark \\ \hline
Societe Generale & France & 516 & 1.703 & gle.fr & gle.fr& \checkmark \\ 
Royal Bank of Scotland Group & UK & 356 & 1.703 & rbs.gb & rbs.ln & \xmark \\ 
Sumitomo Mitsui Financial Group & Japan & 643 & 1.567 & smtm.jp & 8316.to& \checkmark \\ 
Banco Santander & Spain & 1030 & 1.538 & san.es & san.mc& \checkmark \\ 
Wells Fargo & US & 2430 & 1.527 & wfc.us & wfc & \checkmark\\ \hline
ING Groep & Netherland & 557 & 1.490 & inga.nl & inga.ae& \checkmark \\ 
Lloyds Banking Group & UK & 961 & 1.403 & lloy.gb & lloy.ln& \checkmark \\ 
Unicredit & Italy & 477 & 1.166 & ucg.it & ucg.mi& \checkmark \\ 
UBS & Switzerland & 802 & 1.138 & ubsn.ch & ubsn.vx& \checkmark \\ 
Credit Suisse Group & Switzerland & 503 & 983 & csgn.ch & csgn.vx& \checkmark \\ \hline
Goldman Sachs Group & US & 742 & 912 & gs.us & gs& \checkmark \\ 
Nordea Bank & Sweden & 556 & 870 & nor.se & ndasek.sk& \checkmark \\ 
Intesa Sanpaolo & Italy & 458 & 864 & isp.it & isp.mi& \checkmark \\ 
Morgan Stanley & US & 577 & 833 & ms.us & ms& \checkmark \\ 
Toronto-Dominion Bank & Canada & 827 & 827 & td.ca & td.t & \checkmark\\ \hline
Royal Bank of Canada & Canada & 935 & 825 & ry.ca & ry.t & \checkmark\\ 
Banco Bilbao Vizcaya Argentaria & Spain & 708 & 803 & bbva.es & bbva.mc& \checkmark \\ 
Commerzbank & Germany & 206 & 759 & cbk.de & cbk.xe& \checkmark \\ 
National Australia Bank & Australia & 724 & 755 & nab.au & nab.au& \checkmark \\ 
Bank of Nova Scotia & Canada & 698 & 713 & bns.ca & bns.t& \checkmark \\ \hline
Commonwealth Bank of Australia & Australia & 1100 & 688 & cba.au & cba.au & \checkmark\\ 
Standard Chartered & UK & 524 & 674 & stan.gb & stan.ln & \checkmark\\ 
China Merchants Bank & China & 358 & 664 & cmb.cn & 600036.sh & \xmark\\ 
Australia and New Zealand Banking Group & Australia & 776 & 656 & anz.au & anz.au& \xmark \\ 
Westpac Banking & Australia & 918 & 650 & wbc.au & wbc.au& \checkmark \\ \hline
Shanghai Pudong Development Bank & China & 295 & 608 & shgp.cn & 600000.sh & \xmark \\ 
Danske Bank & Denmark & 256 & 597 & dan.dk & danske.ko & \checkmark\\ 
Sberbank Rossii & Russia & 594 & 552 & sber.ru & sber.mz & \checkmark\\ 
China Minsheng Banking Corp & China & 297 & 533 & cmb.cn & 600016.sh & \checkmark\\ 
Bank of Montreal & Canada & 419 & 515 & bmo.ca & bmo.t & \checkmark\\ \hline
Itau Unibanco Holding & Brazil & 332 & 435 & itub4.br & itub4.br& \checkmark \\ 
Resona Holdings & Japan & 122 & 434 & rsnh.jp & 8308.to& \checkmark \\ 
Nomura Holdings & Japan & 256 & 422 & nmrh.jp & 8604.to& \checkmark \\ 
Sumitomo Mitsui Trust Holdings & Japan & 184 & 406 & smtm.jp & 8309.to& \checkmark \\ 
State Bank of India & India & 165 & 400 & sbin.in & sbin.in& \checkmark \\ \hline
DNB ASA & Norway & 289 & 396 & dnb.no & dnb.os& \checkmark \\ 
Svenska Handelsbanken & Sweden & 309 & 388 & shba.se & shba.sk& \checkmark \\ 
Skandinaviska Enskilda Banken & Sweden & 291 & 387 & seba.se & seba.sk & \checkmark\\ 
Canadian Bank of Commerce & Canada & 324 & 382 & cm.ca & cm.t & \checkmark\\ 
Bank of New York Mellon & US & 363 & 374 & bk.us & bk.us& \checkmark \\ \hline
U.S. Bancorp & US & 745 & 364 & usb.us & usb& \checkmark \\ 
Banco Bradesco & Brazil & 235 & 355 & bbdc4.br & bbdc4.br& \checkmark \\ 
KBC Groupe & Belgium & 260 & 333 & kbc.be & kbc.bt& \checkmark \\ 
PNC Financial Services Group & US & 435 & 320 & pnc.us & pnc.us& \checkmark \\ 
DBS Group Holdings & Singapore & 320 & 318 & d05.sg & d05.sg & \checkmark\\ \hline
Ping An Bank & China & 190 & 313 & pab.cn & 000001.sz& \checkmark \\ 
Woori Finance Holdings & Korea & 84 & 309 & wrfh.kr & 053000.se& \xmark \\ 
Dexia & Belgium & 1 & 307 & dexb.be & dexb.bt & \xmark\\ 
Capital One Financial & US & 415 & 297 & cof.us & cof& \checkmark \\ 
Shinhan Financial Group & Korea & 188 & 295 & shf.kr & 055550.se & \checkmark\\ \hline
Swedbank & Sweden & 308 & 284 & swe.se & sweda.sk& \checkmark \\ 
Hua Xia Bank & China & 124 & 276 & hxb.cn & 600015.sh& \checkmark \\ 
Erste Group Bank & Austria & 168 & 276 & ebs.at & ebs.vi& \checkmark \\ 
Banca Monte dei Paschi di Siena & Italy & 29 & 275 & bmps.it & bmps.mi& \checkmark \\ 
State Street Corporation & US & 30 & 243 & stt.us & stt.us& \checkmark \\ \hline
Banco de Sabadell & Spain & 131 & 225 & sab.es & sab.mc& \checkmark \\ 
United Overseas Bank & Singapore & 251 & 225 & uob.sg & u11.sg& \checkmark \\ 
Banco Popular Espanol & Spain & 13 & 204 & pop.es & pop.mc& \xmark \\ 
Industrial Bank of Korea & Korea & 66 & 193 & ibk.kr & 024110.se & \checkmark\\ 
BB\&T Corp & US & 266 & 183 & bbt.us & bbt& \xmark \\ \hline
Bank of Ireland & Ireland & 146 & 182 & bir.ie & bir.db& \checkmark \\ 
National Bank of Canada & Canada & 131 & 180 & na.ca & na.t& \checkmark \\ 
SunTrust Banks & US & 203 & 175 & sti.us & sti.us& \checkmark \\ 
Banco Popolare & Italy & 36 & 174 & bp.it & bp.mi& \xmark \\ 
Malayan Banking Berhad & Malaysia & 263 & 171 & may.my & maybank.ku& \checkmark \\ \hline
Allied Irish Banks & Ireland & 999 & 162 & aib.ie & aib.db& \xmark \\ 
Standard Bank Group & South Africa & 177 & 161 & sbk.za & sbk.jo& \checkmark \\ 
American Express & US & 947 & 153 & axp & axp & \checkmark\\ 
National Bank of Greece & Greece & 121 & 153 & ete.gr & ete.at& \checkmark \\ 
Macquarie Group & Australia & 160 & 143 & mqg.au & mqg.au& \checkmark \\ \hline
Fukuoka Financial Group & Japan & 33 & 137 & ffg.jp & 8354.to& \checkmark \\ 
Bank Of Yokohama & Japan & 63 & 134 & boy.jp & 8332.to & \xmark \\ 
Pohjola Bank & Finland & 58 & 132 & poh.fi & poh1s.he& \xmark \\ 
Fifth Third Bancorp & US & 185 & 130 & fitb.us & fitb.us& \checkmark \\ 
Regions Financial & US & 143 & 117 & rf.us & rf.us & \checkmark\\ \hline
Chiba Bank & Japan & 52 & 117 & cbb.jp & 8331.to& \checkmark \\ 
Unipol Gruppo Finanziario & Italy & 28 & 116 & uni.it & uni.mi& \checkmark \\ 
Banco Comercial Portugues & Portugal & 51 & 113 & bcp.pr & bcp.lb & \checkmark\\ 
CIMB Group Holdings & Malaysia & 163 & 113 & cimb.my & cimb.ku & \checkmark\\ 
Bank of Baroda & India & 37 & 113 & bob.in & bankbaroda.in & \checkmark\\ \hline
Turkiye Is Bankasi & Turkey & 89 & 112 & isctr.tr & isctr.is& \checkmark \\ 
Banco Espirito Santo & Portugal & 71 & 111 & bes.pr & bes.lb & \xmark\\ 
Hokuhoku Financial Group  & Japan & 25 & 108 & hkf.jp & 8377.to & \checkmark\\ 
Shizuoka Bank & Japan & 61 & 104 & shzb.jp & 8355.to & \checkmark\\ 
Mediobanca Banca di Credito Finanziario & Italy & 85 &  95 & mb.it & mb.mi& \checkmark \\ \hline
Yamaguchi Financial Group & Japan & 23 &  93 & yfg.jp & 8418.to   & \checkmark         
\label{Bank List}
\end{longtable}
\end{center}
}

\end{appendix}

\end{document}